\documentclass[aps,pra,reprint,nofootinbib,superscriptaddress]{revtex4-2}

\usepackage[utf8]{inputenc}
\usepackage[T1]{fontenc}

\usepackage{aas_macros}

\usepackage{mathtools,amssymb,tensor}
\usepackage{mathrsfs, stmaryrd}

\usepackage{lmodern,microtype}
\allowdisplaybreaks

\usepackage[dvipsnames]{xcolor}
\usepackage[colorlinks=true,allcolors=Blue]{hyperref}
\usepackage[hyphenbreaks]{breakurl}
\usepackage[capitalize]{cleveref}

\let\t\tensor
\let\p\partial
\let\bs\boldsymbol
\def\transpose{\mathrm{T}}
\def\dd{\mathrm{d}} 
\def\FWD{{\mathscr D}} 
\def\half{\tfrac{1}{2}}

\def\gopt{{\tilde g}} 
\def\frame{\mathfrak e}

\def\jones{{\mathrm J}}
\def\jonesV{\bs{\mathrm J}}
\def\Qq{\bs q}
\def\Mm{{\mathbf M}}
\def\inhom{\Sigma}

\def\tangent{T}     
\def\normal{\nu}    
\def\uNormal{N}     
\def\binormal{B}    
\def\torsion{\tau}
\def\curvature{\kappa}

\newcommand{\jump}[1]{\llbracket #1 \rrbracket} 
\newcommand{\op}[1]{{\mathrm #1}}
\newcommand{\Op}[1]{{\underline{\mathrm #1}}}

\usepackage{siunitx}

\begin{document}
\title{Polarization transport in optical fibers beyond Rytov’s law}

\author{Thomas B. Mieling}
\email{thomas.mieling@univie.ac.at}
\thanks{ORCID:~\href{https://orcid.org/0000-0002-6905-0183}{\texttt{0000-0002-6905-0183}}}
\affiliation{University of Vienna, Faculty of Physics, Vienna Doctoral School in Physics (VDSP), Vienna Center for Quantum Science and Technology (VCQ) and Research platform TURIS, Boltzmanngasse~5, 1090 Vienna, Austria}

\author{Marius A. Oancea}
\email{marius.oancea@univie.ac.at}
\thanks{\hspace{0.45em}ORCID:~\href{https://orcid.org/0000-0002-1242-4041}{\texttt{0000-0002-1242-4041}}}
\affiliation{University of Vienna, Faculty of Physics, Boltzmanngasse 5, 1090 Vienna, Austria}

\begin{abstract}
We consider the propagation of light in arbitrarily curved step-index optical fibers. Using a multiple-scales approximation scheme, set up in Fermi normal coordinates, the full vectorial Maxwell equations are solved in a perturbative manner. At leading order, this provides a rigorous derivation of Rytov’s law. At next order, we obtain non-trivial dynamics of the electromagnetic field, characterized by two coupling constants, the phase and the polarization curvature moments, which describe the curvature response of the light’s phase and its polarization vector, respectively. The latter can be viewed as an inverse spin Hall effect of light, where the direction of propagation is constrained along the optical fiber and the polarization evolves in a frequency-dependent way.
\end{abstract}

\maketitle

\section{Introduction}

Optical fibers are widely used as waveguides for electromagnetic fields, providing a controlled way of transporting light along a given path. They play a fundamental role in many engineering applications, such as telecommunications \cite{senior2009optical} and sensors \cite{LEE200357}. Furthermore, optical fibers play a central role in many areas of physics, such as metrology \cite{doi:10.1126/science.1218442}, photonics \cite{Nayak_2018}, quantum information \cite{Flamini_2019} and quantum computation \cite{PhysRevLett.96.010503,cacciapuoti2019quantum,9023997}, and are planned to be used in future experiments that probe the interplay of quantum optics and Einstein’s theory of gravity \cite{2017NJPh...19c3028H,2018CQGra..35x4001B,2020CQGra..37v5001M,2022PhRvA.106f3511M}.

The propagation of electromagnetic waves in optical fibers is generally described by Maxwell’s equations. For straight optical fibers with circular cross sections, Maxwell’s equations can be solved explicitly using certain mode decompositions adapted to the symmetry of the problem \cite{Liu_2005,Davis_2014,okamoto2021fundamentals}. 
However, in real-world applications, optical fibers are generally twisted and bent in various ways, leading to optical losses, non-trivial polarization dynamics, and coupling between different modes. In this case, exact solutions to Maxwell’s equations are no longer available, which makes it necessary to use approximation schemes.

\begin{figure}[t]
	\centering
	\includegraphics[width=0.8\columnwidth]{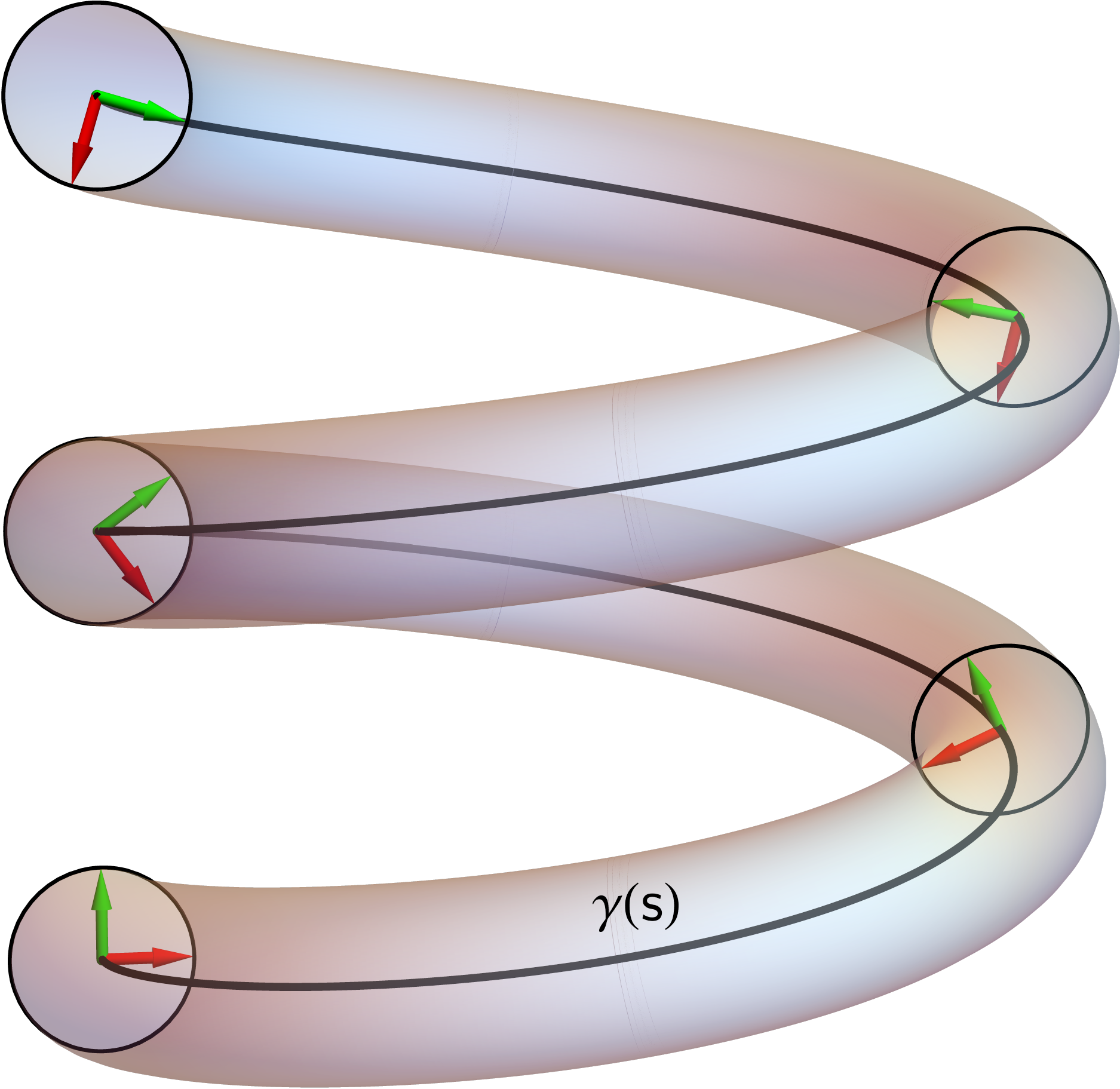}
	\caption{Illustration of Rytov’s law in an optical fiber. The two vectors $e_1$ (red) and $e_2$ (green), orthogonal to the tangent vector $T = \gamma'$ (not shown), are Fermi--Walker transported along the curve $\gamma$, and they can be viewed as linear polarization vectors aligned with the electric field lines of an electromagnetic wave with horizontal or vertical linear polarization. For example, a linearly polarized electromagnetic wave initially at the lower end of the fiber will experience an counterclockwise rotation of its plane of polarization as it propagates upwards along the helix.}
	\label{fig:fiber}
\end{figure}

A geometrical-optics treatment of light rays following nonplanar curves was already given in 1938 by Rytov \cite{rytov}, and later extended by Vladimirskii \cite{Vladimirskii} (see Ref.~\cite{Vinitskii_1990} for a more recent overview of these results). In these papers, the authors derived a transport law for the polarization vector (defined as a unit vector pointing along the direction of the electric field) along the nonplanar curve followed by light rays in inhomogeneous media. This is known as Rytov’s law, which states that the polarization vector is Fermi--Walker transported along the curve \cite[Ch.~6.1]{Chruscinski2012}. In optical fibers, Rytov’s law was experimentally observed by Ross \cite{Ross1984}, followed by a similar experiment by Tomita and Chiao \cite{1986PhRvL..57..937T} (see also Ref.~\cite{PhysRevLett.57.933}). A theoretical discussion describing the geometry of this polarization transport law in optical fibers was given by Haldane in Refs.~\cite{1986OptL...11..730H,1987PhRvL..59.1788H}. We give a simple illustration of Rytov’s law in \cref{fig:fiber}, where we consider a helical fiber, together with the Fermi--Walker transport of two linear polarization vectors.

The standard derivation of Rytov’s law relies on the geometrical optics approximation, which breaks down for electromagnetic waves propagating in single-mode optical fibers (SMFs) both because the wavelength is of the same order of magnitude as the diameter of the fiber and because of the rapid formation of caustics. An extension of Rytov’s law for light propagating in SMFs was given by Berry \cite{1987Natur.326..277B}, where a term proportional to the wavelength was added to the polarization transport law. Similar results were also obtained by Lai \textit{et al.} \cite{2018PhRvA..97c3843L} using different methods.

However, Berry did not provide an explicit derivation of his result from Maxwell’s equations, and the work of Lai \textit{et al.} does not consider the junction conditions of the electromagnetic field at the core-cladding interface, which renders the relation of their result to optical fibers unclear.

In this paper, we perturbatively solve for the full electromagnetic field in an arbitrarily bent step-index optical fiber under the sole assumption that the radius of curvature of the fiber is much larger than the wavelength of light propagating therein. To do so, we erect cylindrical Fermi coordinates around the baseline of the optical fiber, such that the core-cladding interface is at a constant radial coordinate and the effects of bending are fully encoded in a single component of the metric tensor. The assumption of weak bending then allows setting up a perturbative scheme based on a multiple-scales method, where the perturbation parameter is given by the ratio of the optical wavelength to the fiber’s radius of curvature. We mainly focus on the single-mode regime, but we shall also discuss briefly how our method can be applied in certain multi-mode regimes.

At first order in this perturbative expansion, we recover Rytov's law, which means that the Jones vector is Fermi--Walker transported along the fiber. At next order, we obtain corrections to this transport law, describing non-trivial dynamics of the electromagnetic phase and polarization due to the bending of the fiber.
This correction to the polarization dynamics can be interpreted as an inverse spin Hall effect of light: whereas in the (direct) spin Hall effect of light \cite{Onoda2004,Duval2006,Hosten2008,Bliokh2008,Bliokh2009,SOI_review} the polarization influences the trajectory of light, here the trajectory is prescribed by the optical fiber, which then influences the light polarization. Similar inverse spin Hall effects of light have also been reported in other physical systems \cite{10.1038/ncomms6327,nayak2022momentum}.

This paper is structured as follows. In \cref{s:geometry}, we introduce the Fermi coordinate system used in the subsequent calculation and describe the relation of the metric tensor components (in this coordinate system) to the curvature and torsion of the fiber. \Cref{s:maxwell} describes the perturbative ansatz for the electromagnetic potential (using an adapted orthonormal frame along the fiber and the aforementioned multiple-scales approach), and formulates the perturbative equations and their solutions. The expressions that arise there require numerical quadrature. In \cref{s:results} we present numerical results for the polarization moments, which describe the coupling of the electromagnetic phase and polarization to the bending of the fiber, and illustrate the polarization transport for the simple case of a helical fiber.
Finally, in \cref{s:discussion} we compare with previous results obtained by other techniques and provide an outlook on potential future applications of our methods.

\section{Geometric Setup}
\label{s:geometry}

Let $\gamma$ be a curve in three-dimensional Euclidean space representing the baseline of an SMF.
In the vicinity of this curve, one can erect a Fermi coordinate system $(t, x, y, s)$, where $t$ is time, $s$ is the arc length of $\gamma$, and $x$ and $y$ are transverse coordinates. More details are given in \cref{s:Fermi coordinates}.
In this coordinate system, the Minkowski metric takes the form
\begin{equation}
	g
		=
		- c^2 \dd t^2
		+ \dd x^2
		+ \dd y^2
		+ [ 1 - \t\normal{_x}(s) x - \t\normal{_y}(s) y]^2 \dd s^2\,,
\end{equation}
where $c$ is the speed of light in vacuum and $\t\normal{_i}$ are the components of the normal vector of the curve in the Fermi coordinate system.

Henceforward, we assume that the fiber is only weakly bent in the following sense.
First, we require the radius of curvature $1/\curvature = 1/\sqrt{\t*\normal{_x^2} + \t*\normal{_y^2}}$ to be much larger than the SMF’s core radius $\varrho$, i.e.\ $\varrho \kappa \ll 1$. In practice, this requirement is well satisfied, since typical SMFs, bent on the scale of centimeters, have $\varrho \curvature \sim \si{\micro\meter}/\si{\centi\meter} = \num{e-4}$.
Second, we require the normal vector $\t\normal{_i}$ to be not only small in norm (of order $\curvature \ll 1/\varrho$), but also to vary slowly when measured in units of the radius $\varrho$. This means that $\t\normal{_i}$ can be expressed in the form
\begin{equation}
	\t\normal{_i}(s)
		= \varepsilon \varrho^{-1} \t b{_i}(\varepsilon \varrho^{-1} s)\,,
\end{equation}
where $\varepsilon$ is a small parameter and $\t b{_i}$ are dimensionless functions whose derivatives are at most of the same order of magnitude as the functions themselves.
In practice, since SMF modes have wavelengths comparable to the core radius $\varrho$, this means that the normal vector is almost constant over an optical wavelength and changes significantly only over a length scale of $1 / \varepsilon \gg 1$ wavelengths.

To simplify the notation, all subsequent calculations will be carried out in nondimensionalized cylindrical coordinates $(\tilde t, r, \vartheta, \sigma)$, defined by
\begin{align}
	t &= \varrho  \tilde t / c\,,
	&
	x &= \varrho r \cos \vartheta\,,
	&
	y &= \varrho r \sin \vartheta\,,
	&
	s &= \varrho \sigma\,.
\end{align}
%
In this coordinate system, the Minkowski metric takes the form 
\begin{equation}
	\label{eq:metric physical}
	g =
		\varrho^2 [
		- \dd{\tilde t}^2
		+ \dd r^2
		+ r^2 \dd \vartheta^2
		+ (1 - \varpi)^2 \dd \sigma^2
		]\,,
\end{equation}
where
\begin{equation}
	\varpi
		= \frac{\varepsilon r}{\sqrt 2} \left[
			\t b{_+}(\varsigma) e^{+ i \vartheta}
			+ \t b{_-}(\varsigma) e^{- i \vartheta}
		\right]\,,
\end{equation}
with $\varsigma = \varepsilon \sigma$ and $\t b{_\pm}$ being the complex bending functions
\begin{equation}
	\t b{_\pm} = \tfrac{1}{\sqrt 2} (\t b{_x} \mp i \t b{_y})\,.
\end{equation}
These complex functions are related to the fiber’s curvature $\kappa$ and torsion $\tau$ by
\begin{equation}
	\varepsilon \t b{_\pm} = \frac{\varrho \kappa}{\sqrt 2} \exp\left[
		\mp i \left( \int_0^s  \tau(s') \, \dd s' + \text{const}.\right)
	\right]\,,
\end{equation}
where the arbitrary constant describes the freedom to rotate the coordinate system rigidly around the $r = 0$ axis, cf.~\cref{eq:normal complex components}.

\section{Maxwell’s Equations}
\label{s:maxwell}

The electromagnetic field in optical fibers satisfies the source-free Maxwell equations
\begin{align}
	\operatorname{div} \bs B &= 0\,,
	&
	\operatorname{curl} \bs E\; + \t\p{_t} \bs B / c &= 0\,,
	\\
	\operatorname{div} \bs D &= 0\,,
	&
	\operatorname{curl} \bs H - \t\p{_t} \bs D / c &= 0\,,
\end{align}
when using either Gaussian or Lorentz--Heaviside units.
For non-magnetic materials, the constitutive relations are
\begin{align}
	\bs D &= n^2 \bs E\,,
	&
	\bs H &= \bs B\,,
\end{align}
with $n$ denoting the refractive index. For step-index optical fibers, $n$ has the form
\begin{equation}
	n(r)
		= \begin{cases}
			n_1 &   0 < r < 1\,,\\
			n_2 &   1 < r\,,
		\end{cases}
\end{equation}
where the refractive indices of the core and the cladding, $n_1$ and $n_2$, are constants satisfying $n_1 > n_2 \geq 1$.
Here, the operators $\operatorname{div}$ and $\operatorname{curl}$ are those of flat Euclidean space, which are to be expressed in non-Cartesian coordinates when working with the curvilinear Fermi coordinate system.

\subsection{Gauge-Fixed Field Equations}

Here, we work with the electromagnetic potentials $\phi$ and $\bs A$, such that
\begin{align}
	\bs E &= - \operatorname{grad} \phi -  \t\p{_t} \bs A / c\,,
	&
	\bs B &= \operatorname{curl} \bs A\,.
\end{align}
Imposing an appropriate generalization of the Lorenz gauge to linear isotropic dielectrics, as in Ref.~\cite{2022PhRvA.106f3511M}, the field equations reduce to
\begin{align}
	\label{eq:field equation abstract}
	n^2 \t*\p{_t^2} \phi &= c^2 \Delta \phi\,,
	&
	n^2 \t*\p{_t^2} \bs A &= c^2 \Delta \bs A\,,
\end{align}
wherever $n$ is constant. Here, $\Delta$ denotes the Laplacian (more precisely: the scalar Laplace--Beltrami operator when acting on $\phi$, for the vector Laplace--Beltrami operator when acting on $\bs A$) of Euclidean space, which we express in curvilinear coordinates below.

Equation \eqref{eq:field equation abstract} can also be written in generally covariant form. In regions of constant $n$, the electromagnetic four-potential $A = - c \phi \dd t + \t A{_i} \t{\dd x}{^i}$ satisfies the wave equation
\begin{equation}
	\label{eq:field equation abstract - covariant}
	\t\gopt{^a^b} \t\nabla{_a} \t\nabla{_b} \t A{_c} = 0\,,
\end{equation}
(in abstract index notation) where $\t\nabla{_a}$ is the Levi-Civita derivative of the metric $\t g{_a_b}$ given in \cref{eq:metric physical} and $\t\gopt{^a^b}$ is the inverse of Gordon’s optical metric $\t\gopt{_a_b}$ \cite{1923AnP...377..421G}, which here takes the form
\begin{equation}
	\label{eq:metric optical}
	\gopt =
		\varrho^2 [
		- n^{-2} \dd{\tilde t}^2
		+ \dd r^2
		+ r^2 \dd \vartheta^2
		+ (1 - \varpi)^2 \dd \sigma^2
		]\,.
\end{equation}

To decouple the wave equations as much as possible, we use an adapted frame $\t\frame{_\mu}$ and a coframe $\t\frame{^\mu}$:
\begin{align}
	\t\frame{_0} &= \t\p{_{\tilde t}}\,,
	&
	\t\frame{_\parallel} &= \frac{1}{1 - \varpi}\, \t\p{_\sigma}\,,
	&
	\t\frame{_\pm} &= \tfrac{1}{\sqrt 2}[\t\p{_r} \pm \tfrac{i}{r} \t\p{_\vartheta}]\,,
	\\
	\t\frame{^0} &= \dd {\tilde t}\,,
	&
	\t\frame{^\parallel} &= (1 - \varpi) \dd \sigma\,,
	&
	\t\frame{^\pm} &= \tfrac{1}{\sqrt 2}[\dd r \mp i r \dd\vartheta]\,,
\end{align}
with respect to which the electromagnetic four-potential can be decomposed as
\begin{equation}
	A
		= \t A{_\mu} \t\frame{^\mu}
		\equiv
		\t A{_0} \t\frame{^0}
		+ \t A{_\parallel} \t\frame{^\parallel}
		+ \t A{_+} \t\frame{^+}
		+ \t A{_-} \t\frame{^-}\,.
\end{equation}
To express the junction conditions at the core-cladding interface $r = 1$ (derived in Ref.~\cite{2022PhRvA.106f3511M}) in a form suitable for the following calculations, define the column vector
\begin{widetext}
\begin{equation}
	\label{eq:discontinuity vector}
	\bs\Psi[A]
		= \begin{pmatrix}
			\jump{\t A{_0}} &
			\jump{\t A{_\parallel}} &
			\jump{\t A{_+}} &
			\jump{\t A{_-}} &
			\jump{\t*A{_\parallel^\prime}} &
			\jump{\t*A{_+^\prime} - \t*A{_-^\prime}} &
			\jump{n^2 (\t*A{_0^\prime} - [\t{\dot A}{_+} + \t{\dot A}{_-}]/\sqrt 2)} &
			\jump{n^2 \t{\dot A}{_0} - [\t*A{_+^\prime} + \t*A{_-^\prime}]/\sqrt 2}
		\end{pmatrix}^\transpose\,,
\end{equation}
\end{widetext}
where dots and primes denote partial differentiation with respect to $\tilde t$ and $r$, respectively, and, for any function $\varphi$, the expression $\jump{\varphi}$ denotes the discontinuity of $\varphi$ at the interface:
\begin{equation}
	\jump{\varphi}
		= \left( \lim_{r \nearrow 1} \varphi \right)
		- \left( \lim_{r \searrow 1} \varphi \right)\,.
\end{equation}
The junction conditions can then be expressed as ${\bs\Psi[A] = 0}$, which means that all field expressions entering \cref{eq:discontinuity vector} must be continuous.

\begin{figure*}
	\centering
	\includegraphics[width=0.9\columnwidth]{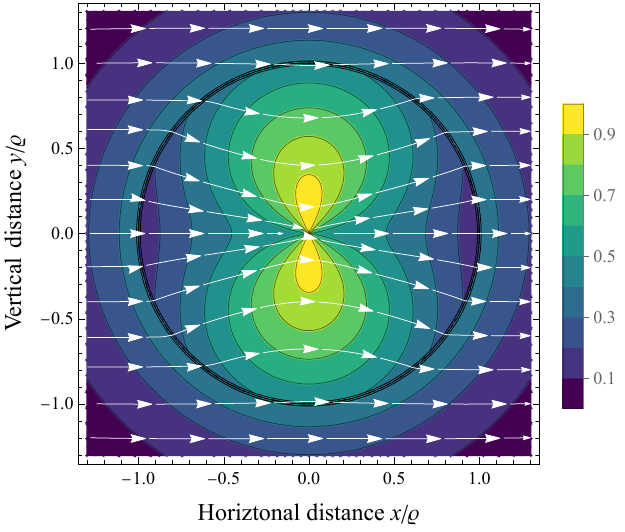}
	\hspace{0.2\columnwidth}
	\includegraphics[width=0.9\columnwidth]{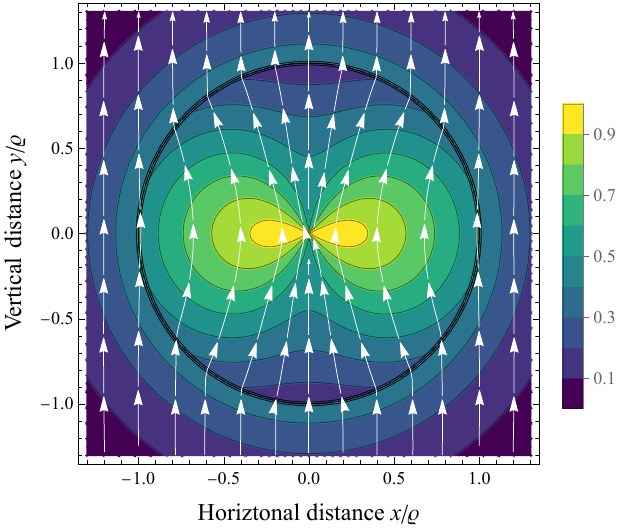}
	\caption{%
		Transverse electric field profiles of linearly polarized fiber modes (LP modes).
		Horizontal polarization corresponds to the Jones vector $\t\jones{_\pm} = 1 / \sqrt 2$ (left figure),
		while vertical polarization corresponds to $\t\jones{_\pm} = \pm i / \sqrt 2$ (right figure).
		For illustration, we have set the refractive indices to $n_1 = 1.46$, $n_2 = 1$, and the normalized frequency $V = (\varrho \omega/c) \sqrt{n_1^2 - n_2^2}$ to $V = 2$. The contours indicate the relative electric field intensity.
	}
	\label{fig:plot:polarization}
\end{figure*}

\subsection{Perturbative system}

In the absence of bending, the fiber modes have constant angular frequency $\omega$ and propagation constant $\beta$, for which we define the dimensionless counterparts
\begin{align}
	\tilde\beta
		&= \varrho \beta\,,
	&
	\tilde\omega
		&= \varrho \omega/c\,.
\end{align}
To set up a perturbative scheme for \cref{eq:field equation abstract} in the presence of bending, we make the following ansatz.
Seeking monochromatic waves of angular frequency $\omega$ and propagation constant $\beta$, we write
\begin{equation}
	\t A{_\mu}
		= \t a{_\mu}(r, \vartheta, \varsigma) e^{i(\tilde\beta \sigma - \tilde\omega \tilde t)}\,,
\end{equation}
where the field’s angular dependence is expanded in a Fourier series:
\begin{equation}
	\t a{_\mu}(r, \vartheta, \varsigma)
		= \sum_{m \in \mathbf Z} \t* a{_\mu^{(m)}}(r, \varsigma) e^{i m \vartheta}\,,
\end{equation}
with the following perturbative expansion
\begin{equation}
	\t* a{_\mu^{(m)}}(r, \varsigma)
		= \sum_{k = ||m|-1|}^\infty \varepsilon^k \t* a{_\mu^{(m,k)}}(r, \varsigma)\,.
\end{equation}
Here, $\sigma$ and $\varsigma = \varepsilon \sigma$ are to be considered as “independent variables” in the sense of the multiple-scales method \cite{1978amms.book.....B}.
The precise form of the last expansion is motivated as follows. In the unperturbed problem, SMFs allow for modes with $m = +1$ and $m = -1$ only \cite{Liu_2005,Davis_2014}. Hence, we start the series expansion of these Fourier modes at order $\varepsilon^0$. The bending terms couple neighboring Fourier components, so the series expansions of $m = 0$ and $m = \pm 2$ start at order $\varepsilon^1$, while Fourier modes with $m = \pm 3$ are at most of order $\varepsilon^2$, and so on.

We insert this ansatz into \cref{eq:field equation abstract} and consider terms of various powers of $\varepsilon$ separately.

\subsubsection{Unperturbed system}

At order $\varepsilon^0$, one obtains
\begin{equation}
	{\Op H}_{\pm 1} \t*a{_\mu^{(\pm 1, 0)}} = 0\,,
\end{equation}
where the Helmholtz operator ${\Op H}_m$ acts on the fields
\begin{equation}
	\label{eq:vector a mu}
	\t a{_\mu} = (
		\t a{_0},
		\t a{_\parallel},
		\t a{_+},
		\t a{_-}
	)^\transpose
\end{equation}
as follows:
\begin{align}
	{\Op H}_{m} \t a{_\mu}
		&= ( {\op H}_{m} \t a{_0}, {\op H}_{m} \t a{_\parallel}, {\op H}_{m+1} \t a{_+}, {\op H}_{m-1} \t a{_-} )^\transpose\,,
	\\
	\shortintertext{with}
	{\op H}_{m}
		&= \frac{\p^2}{\p r^2} + \frac{1}{r} \frac{\p}{\p r} - \frac{m^2}{r^2} + n^2 \tilde\omega^2 - \tilde\beta^2\,.
\end{align}
The solution which is regular on the axis $r = 0$ and decays for $r \gg 1$ is given in terms of Bessel functions as
\begin{subequations}\label{eq:solution unperturbed 1}
\begin{align}
	\t*a{_0^{(m, 0)}} &= \t*f{^{(m,\phantom+0)}}(\t*q{^{(m,0)}_{\,0,1}}, \t*q{^{(m,0)}_{\,0,2}}, r)\,,\\
	\t*a{_\parallel^{(m, 0)}} &= \t*f{^{(m,\phantom+0)}}(\t*q{^{(m,0)}_{\,\parallel,1}}, \t*q{^{(m,0)}_{\,\parallel,2}}, r)\,,\\
	\t*a{_+^{(m, 0)}} &= \t*f{^{(m,+1)}}(\t*q{^{(m,0)}_{\,+,1}}, \t*q{^{(m,0)}_{\,+,2}}, r)\,,\\
	\t*a{_-^{(m, 0)}} &= \t*f{^{(m,-1)}}(\t*q{^{(m,0)}_{\,-,1}}, \t*q{^{(m,0)}_{\,-,2}}, r)\,.
\end{align}
\end{subequations}
The coefficients $\t*q{^{(m,k)}_{\mu,l}}$ may depend on $\varsigma$, and the functions $f^{(m)}$ are defined in terms of Bessel functions of the first kind, $J_m$, and modified Bessel functions of the second kind, $K_m$, as
\begin{equation}
	f^{(m, m')}(q_1, q_2, r)
		= \begin{cases}
			q_1 \frac{J_{m+m'}(U r)}{J_m(U)}    &   r < 1\,,\\
			q_2 \frac{K_{m+m'}(W r)}{K_m(W)}    &   r > 1\,,
		\end{cases}
\end{equation}
with
\begin{align}
	U &= \sqrt{+ n_1^2 \tilde\omega^2 - \tilde\beta^2}\,,
	&
	W &= \sqrt{- n_2^2 \tilde\omega^2 + \tilde\beta^2}\,.
\end{align}
For conciseness, we shall abbreviate \cref{eq:solution unperturbed 1} as
\begin{equation}
	\t*a{_\mu^{(m, 0)}}
		= \t*f{_\mu^{(m)}}(\Qq^{(m,0)})\,,
\end{equation}
where $\Qq^{(m,0)}$ is a column vector containing all the parameters $q$ that occur in \cref{eq:solution unperturbed 1}.

The matching conditions at the core-cladding interface $r = 1$ can then be written in matrix form
\begin{equation}
	\label{eq:continuity order 0}
	\bs\Psi[\t*a{_\mu^{(\pm 1, 0)}}]
		\equiv \t\Mm{_{\pm 1}} \cdot \t\Qq{^{(\pm 1, 0)}}
		= 0\,,
\end{equation}
where $\Mm_m$ is a complex $8 \times 8$ matrix, given explicitly by Eq.~(58) in Ref.~\cite{2022PhRvA.106f3511M}, the precise details of which are not important for our present considerations.

For \cref{eq:continuity order 0} to admit a non-trivial solution, the determinant of $\t\Mm{_{\pm 1}}$ must vanish (one can show that the two determinants for $m = +1$ and $m = -1$ are proportional, so that they must vanish simultaneously). This determines the standard dispersion relation between $\beta$, $\omega$, $\varrho$ and the refractive indices $n_1, n_2$ \cite{Liu_2005,Davis_2014}.

If the dispersion relation is satisfied (SMFs admit only a single solution for $\beta$ in terms of $\omega$), the matrices $\t\Mm{_{\pm 1}}$ possess a one-dimensional kernel and co-kernel.
Setting $\t{\bs{\hat q}}{_{\pm}}$ to be any vector spanning the kernel of $\t\Mm{_{\pm 1}}$ (we take $\t{\bs{\hat q}}{_{\pm}}$ to be $\varsigma$-independent), the general solution to \cref{eq:continuity order 0} can be written as
\begin{equation}
	\t\Qq{^{(\pm 1, 0)}}
		= \t*\jones{_\pm^{(0)}}(\varsigma) \t{\bs{\hat q}}{_{\pm}}\,,
\end{equation}
where $\t*\jones{_\pm^{(0)}}$ can be interpreted as complex components of a ($\varsigma$-dependent) Jones vector, as the solution with $m = +1$ corresponds to right circular polarization, while $m = -1$ describes left circular polarization.
Here, we choose the norms and global phases of the vectors $\t{\bs{\hat q}}{_{\pm}}$ such that ${\t*\jones{_\pm^{(0)}} = 1/\sqrt 2}$ corresponds to horizontal polarization and ${\t*\jones{_\pm^{(0)}} = \pm i / \sqrt 2}$ corresponds to vertical polarization; see \cref{fig:plot:polarization}.
To determine the dependence of $\t*\jones{_\pm^{(0)}}$ on $\varsigma$, we must consider the equations at next order in the perturbative expansion.

\subsubsection{First order perturbations}

At order $\varepsilon^1$, one obtains differential equations of the form
\begin{subequations}
\label{eq:perturbative system 1}
\begin{align}
	{\Op H}_{0\phantom+} \t*a{_\mu^{\phantom+(0,1)}}
		&= \t*\inhom{_\mu^{\phantom+(0,1)}}\,,
		\\
	{\Op H}_{\pm 1} \t*a{_\mu^{(\pm 1,1)}}
		&= \t*\inhom{_\mu^{(\pm 1,1)}}\,,
		\\
	{\Op H}_{\pm 2} \t*a{_\mu^{(\pm 2,1)}}
		&= \t*\inhom{_\mu^{(\pm 2,1)}}\,,
\end{align}
\end{subequations}
where the inhomogeneities $\t*\inhom{_\mu^{(m,1)}}$, given explicitly in \cref{s:inhomogeneities}, depend on the fields at order $\varepsilon^0$, their first $\varsigma$-derivatives, as well as on the bending functions $\t b{_\pm}$.

These equations can be solved using Green’s functions for the Helmholtz operator ${\Op H}_m$ \cite[Ch.~1.16]{korenev2002bessel}.
Setting
\begin{align}
	{\Op G}_{m} \t a{_\mu}
		&= ( {\op G}_{m} \t a{_0}, {\op G}_{m} \t a{_\parallel}, {\op G}_{m+1} \t a{_+}, {\op G}_{m-1} \t a{_-} )^\transpose\,,
	\\
	\shortintertext{with}
	 {\op G}_{m} \varphi (r)
		&= \begin{cases}
			 {\op G}_{m}^\text{(1)} \varphi (r)
				& r < 1\,,
				\\
			 {\op G}_{m}^\text{(2)} \varphi (r)
				& r > 1\,,
		\end{cases}
	\\
	\begin{split}
	 {\op G}_{m}^\text{(1)} \varphi (r)
		&= \frac{\pi}{2} Y_m(U r) \int_0^r J_m(U r') \varphi(r') r'\,\dd r'
			\\&\quad 
			+ \frac{\pi}{2} J_m(U r) \int_r^1 Y_m(U r') \varphi(r') r'\,\dd r'
		\,,
	\end{split}
	\\
	\begin{split}
	 {\op G}_{m}^\text{(2)} \varphi (r)
		&= - I_m(W r) \int_r^\infty K_m(W r') \varphi(r') r'\,\dd r'
			\\&\quad
			- K_m(W r) \int_1^r I_m(W r') \varphi(r') r'\,\dd r'
		\,,
	\end{split}
\end{align}
where $Y_m$ and $I_m$ denote Bessel functions of the second kind and modified Bessel functions of the first kind, respectively, the general solution can be written in the form
\begin{subequations}
\label{eq:perturbation 1}
\begin{align}
	\t*a{_\mu^{\phantom+(0,1)}}
		&= {\Op G}_{0\phantom+} \t*\inhom{_\mu^{\phantom{\pm}(0,1)}}
		+ \t*f{_\mu^{\phantom+(0)}}(\t\Qq{^{\phantom{\pm}(0,1)}})\,,
		\\
	\t*a{_\mu^{(\pm 1,1)}}
		&= {\Op G}_{\pm 1} \t*\inhom{_\mu^{(\pm 1,1)}}
		+ \t*f{_\mu^{(\pm 1)}}(\t\Qq{^{(\pm 1,1)}})\,,
		\\
	\t*a{_\mu^{(\pm 2,1)}}
		&= {\Op G}_{\pm 2} \t*\inhom{_\mu^{(\pm 2,1)}}
		+ \t*f{_\mu^{(\pm 2)}}(\t\Qq{^{(\pm 2,1)}})\,.
\end{align}
\end{subequations}
The free parameters $\t\Qq{^{(m,1)}}$ are needed to satisfy the junction conditions at the core-cladding interface.
As in the unperturbed case, these matching conditions can be written in matrix form:
\begin{subequations}
\label{eq:continuity 1}
\begin{align}
	\label{eq:continuity 1 0}
	\t\Mm{_{0\phantom+}} \cdot \t\Qq{^{\phantom+(0,1)}}
		&= - \bs\Psi[{\Op G}_{0\phantom+} \t*\inhom{_\mu^{\phantom{\pm}(0,1)}}] \,,
	\\
	\label{eq:continuity 1 1}
	\t\Mm{_{\pm 1}} \cdot \t\Qq{^{(\pm 1,1)}}
		&=- \bs\Psi[{\Op G}_{\pm 1} \t*\inhom{_\mu^{(\pm 1,1)}}] \,,
	\\
	\label{eq:continuity 1 2}
	\t\Mm{_{\pm 2}} \cdot \t\Qq{^{(\pm 2,1)}}
		&= - \bs\Psi[{\Op G}_{\pm 2} \t*\inhom{_\mu^{(\pm 2,1)}}] \,.
\end{align}
\end{subequations}
Since the matrices $\t\Mm{_0}$ and $\t\Mm{_{\pm 2}}$ are non-singular, \cref{eq:continuity 1 0,eq:continuity 1 2} can be solved uniquely for the coefficients $\t\Qq{^{(0,1)}}$ and $\t\Qq{^{(\pm 2,1)}}$.
However, $\t\Mm{_{\pm 1}}$ is singular due to the dispersion relation.
The condition that \cref{eq:continuity 1 1} is solvable is equivalent to the right-hand side of \cref{eq:continuity 1 1} being in the image of the matrix $\t\Mm{_{\pm 1}}$.
Denoting by $\t{\bs \zeta}{_\pm}$ any eight-component row vector that spans the co-kernel of $\t\Mm{_{\pm 1}}$, the solubility condition can be expressed as
\begin{equation}
	\t{\bs \zeta}{_\pm} \cdot \bs\Psi[{\Op G}_{\pm 1} \t*\inhom{_\mu^{(\pm 1,1)}}] = 0\,.
\end{equation}
Expanding all definitions, one obtains equations of the form
\begin{equation}
	\text{(const.)}\, \frac{\dd}{\dd \varsigma} \t*\jones{_\pm^{(0)}}(\varsigma) = 0\,,
\end{equation}
where the multiplicative factor is expressible in terms of integrals of Bessel functions.
Numerically, we find this factor to be non-zero, leading to
\begin{equation}
	\label{eq:jones transport 0}
	\frac{\dd}{\dd \varsigma} \t*\jones{_\pm^{(0)}}(\varsigma) = 0\,.
\end{equation}
As will be explained in more detail below, this equation implies that, to first order in the perturbative expansion, the Jones vector of the SMF is Fermi--Walker transported along the fiber (Rytov’s law).
Non-trivial dynamics arise only at second order.

As \cref{eq:jones transport 0} implies that \cref{eq:continuity 1 1} reduces to ${\t\Mm{_{\pm 1}} \cdot \t\Qq{^{(\pm 1,1)}} = 0}$, it follows that the vectors $\t\Qq{^{(\pm 1,1)}}$ are proportional to the vectors $\t{\bs{\hat q}}{_{\pm}}$ spanning the kernel of $\t\Mm{_{\pm 1}}$ where the proportionality coefficients may depend on $\varsigma$. Hence, we obtain
\begin{equation}
	\t\Qq{^{(\pm 1, 1)}}
		= \t*\jones{_\pm^{(1)}}(\varsigma) \t{\bs{\hat q}}{_{\pm}}\,,
\end{equation}
where the dynamics of the coefficients $\t*\jones{_\pm^{(1)}}$ is to be determined at the next order.

\subsubsection{Second order perturbations}

At order $\varepsilon^2$, one obtains field equations of the form
\begin{subequations}
\label{eq:perturbative system 2}
\begin{align}
	{\Op H}_{0\phantom+} \t*a{_\mu^{\phantom+(0,2)}}
		&= \t*\inhom{_\mu^{\phantom{\pm}(0,2)}}\,,
		\\
	{\Op H}_{\pm 1} \t*a{_\mu^{(\pm 1,2)}}
		&= \t*\inhom{_\mu^{(\pm 1,2)}}\,,
		\\
	{\Op H}_{\pm 2} \t*a{_\mu^{(\pm 2,2)}}
		&= \t*\inhom{_\mu^{(\pm 2,2)}}\,,
	\\
	{\Op H}_{\pm 3} \t*a{_\mu^{(\pm 3,2)}}
		&= \t*\inhom{_\mu^{(\pm 3,2)}}\,,
\end{align}
\end{subequations}
where the inhomogeneities $\t*\inhom{_\mu^{(m,2)}}$, given explicitly in \cref{s:inhomogeneities}, depend on the fields at order $\varepsilon^0$ and $\varepsilon^1$, their $\varsigma$-derivatives, as well as on the bending functions and their derivatives.

These equations can be solved using the Green’s operator defined above:
\begin{subequations}
\label{eq:perturbation 2}
\begin{align}
	\t*a{_\mu^{\phantom+(0,2)}}
		&= {\Op G}_{0\phantom+} \t*\inhom{_\mu^{\phantom\pm(0,2)}}
		+ \t*f{_\mu^{\phantom+(0)}}(\t\Qq{^{\phantom{\pm}(0,2)}})\,,
		\\
	\t*a{_\mu^{(\pm 1,2)}}
		&= {\Op G}_{\pm 1} \t*\inhom{_\mu^{(\pm 1,2)}}
		+ \t*f{_\mu^{(\pm 1)}}(\t\Qq{^{(\pm 1,2)}})\,,
		\\
	\t*a{_\mu^{(\pm 2,2)}}
		&= {\Op G}_{\pm 2} \t*\inhom{_\mu^{(\pm 2,2)}}
		+ \t*f{_\mu^{(\pm 2)}}(\t\Qq{^{(\pm 2,2)}})\,,
		\\
	\t*a{_\mu^{(\pm 3,2)}}
		&= {\Op G}_{\pm 3} \t*\inhom{_\mu^{(\pm 3,2)}}
		+ \t*f{_\mu^{(\pm 3)}}(\t\Qq{^{(\pm 3,2)}})\,,
\end{align}
\end{subequations}
and the interface conditions take the abstract form
\begin{subequations}
\label{eq:continuity 2}
\begin{align}
	\label{eq:continuity 2 0}
	\t\Mm{_{0\phantom+}} \cdot \t\Qq{^{\phantom+(0,2)}}
		&= - \bs\Psi[\Op{G}_{0\phantom+} \t*\inhom{_\mu^{\phantom\pm(0,2)}}] \,,
	\\
	\label{eq:continuity 2 1}
	\t\Mm{_{\pm 1}} \cdot \t\Qq{^{(\pm 1,2)}}
		&=- \bs\Psi[\Op{G}_{\pm 1} \t*\inhom{_\mu^{(\pm 1,2)}}] \,,
	\\
	\label{eq:continuity 2 2}
	\t\Mm{_{\pm 2}} \cdot \t\Qq{^{(\pm 2,2)}}
		&= - \bs\Psi[\Op{G}_{\pm 2} \t*\inhom{_\mu^{(\pm 2,2)}}] \,,
	\\
	\label{eq:continuity 2 3}
	\t\Mm{_{\pm 3}} \cdot \t\Qq{^{(\pm 3,2)}}
		&= - \bs\Psi[\Op{G}_{\pm 3} \t*\inhom{_\mu^{(\pm 3,2)}}] \,.
\end{align}
\end{subequations}
The structure is similar to \cref{eq:continuity 1}.
\Cref{eq:continuity 2 0,eq:continuity 2 2,eq:continuity 2 3} can be solved uniquely for the coefficients $\t\Qq{^{(0,2)}}$, $\t\Qq{^{(\pm 2,2)}}$, and $\t\Qq{^{(\pm 3,2)}}$. However, as the matrices $\t\Mm{_{\pm 1}}$ are singular, the solubility of \cref{eq:continuity 2 1} provides a non-trivial constraint, which is equivalent to
\begin{equation}
	\t{\bs \zeta}{_\pm} \cdot \bs\Psi[{\Op G}_{\pm 1} \t*\inhom{_\mu^{(\pm 1,2)}}] = 0\,.
\end{equation}
Expanding the definitions, this leads to an equation of the form
\begin{equation}
	\label{eq:jones transport 1}
	\frac{\dd}{\dd \varsigma}
	\begin{bmatrix}
		\t*\jones{_+^{(1)}} \\
		\t*\jones{_-^{(1)}}
	\end{bmatrix}
	=
	i \mathbf{\hat B}(\varsigma)
	\begin{bmatrix}
		\t*\jones{_+^{(0)}} \\
		\t*\jones{_-^{(0)}}
	\end{bmatrix}\,,
\end{equation}
where $\mathbf{\hat B}$ is a complex $2 \times 2$ matrix, which we find to be of the general form
\begin{equation}
	\mathbf{\hat B}
		= \begin{bmatrix}
			2 \tilde\eta\, \t b{_+} \t b{_-} &
			\tilde\xi\, \t*b{_+^2}\\
			\tilde\xi\, \t*b{_-^2} &
			2 \tilde\eta\, \t b{_+} \t b{_-}
		\end{bmatrix}\,,
\end{equation}
with $\tilde\xi$ and $\tilde\eta$ being dimensionless constants that depend on the details of the SMF.
(The factors of two on the diagonal were inserted to simplify later calculations using  $2 \t b{_+} \t b{_-} = \varrho^2 \kappa^2 / \varepsilon^2$.)
The matrix $\mathbf{\hat B}$ is Hermitian since the bending functions $\t b{_\pm}$ are each other’s complex conjugates and the parameters $\tilde\xi$ and $\tilde\eta$ are real (this was verified numerically, see \cref{s:results} below).
 
\subsection{Polarization transport}

Combining $\t*\jones{_\pm^{(0)}}$ and $\t*\jones{_\pm^{(1)}}$ into one Jones vector
\begin{align}
	\t\jones{_\pm}
		= \t*\jones{_\pm^{(0)}}
		+ \varepsilon \t*\jones{_\pm^{(1)}}\,,
\end{align}
\cref{eq:jones transport 0,eq:jones transport 1} lead to
\begin{equation} \label{eq:transp_jones}
	\frac{\dd}{\dd s} \bs\jones
		= i \varrho^{-1} \varepsilon^2\, \mathbf{\hat B} \cdot \bs\jones\,,
\end{equation}
where error terms of order $\varepsilon^3$ have been neglected.

Transforming the complex components $\t\jones{_\pm}$ to Cartesian components $\t\jones{_x}$ and $\t\jones{_y}$ according to
\begin{equation}
	\label{eq:jones transport +-}
	\t\jones{_\pm}
		= \tfrac{1}{\sqrt 2}(\t\jones{_x} \mp i \t\jones{_y})\,,
\end{equation}
the evolution equation for the Jones vector $\jonesV$ can be written as
\begin{equation}
	\label{eq:jones transport xy}
	\frac{\dd}{\dd s}\!
	\begin{bmatrix}
		\t\jones{_x} \\
		\t\jones{_y}
	\end{bmatrix}
	=
	i \xi
	\begin{bmatrix}
		\t*\normal{_x^2} - \curvature^2/2    &   \t\normal{_x} \t\normal{_y}\\
		\t\normal{_x} \t\normal{_y}    &   \t*\normal{_y^2}  - \curvature^2/2
	\end{bmatrix}\!
	\begin{bmatrix}
		\t\jones{_x} \\
		\t\jones{_y}
	\end{bmatrix}
	+ i \eta \curvature^2\!
	\begin{bmatrix}
		\t\jones{_x} \\
		\t\jones{_y}
	\end{bmatrix}
	\,,
\end{equation}
where the polarization curvature moment $\xi$ and the phase curvature moment $\eta$ are given by $\xi = \varrho\, \tilde\xi$ and $\eta = \varrho\, \tilde\eta$, respectively. Both curvature moments have dimensions of length. The phase curvature moment $\eta$ contributes to the polarization dynamics by an overall phase, while the polarization curvature moment $\xi$ determines the rate of change of the polarization vector. This can be viewed as an inverse spin Hall effect of light, where the evolution of the state of polarization depends on the wavelength and on the bending of the optical fiber.

Extending the Jones vector $\jonesV$ to a three-dimensional vector by setting $\t\jones{_\parallel} = 0$, this result can be written in covariant vector notation as
\begin{equation}
	\label{eq:jones transport covariant}
	\FWD_s \t\jones{^k}
		= i \xi (\t\normal{^k} \t\normal{_l} - \half \curvature^2 \t*\delta{^k_l}) \t\jones{^l} 
		+ i \eta \curvature^2 \t\jones{^k}\,,
\end{equation}
where $\FWD_s$ is the spatial Fermi--Walker derivative along the fiber’s baseline, defined explicitly in \cref{eq:FW derivative} below.

\section{Results}
\label{s:results}

We have implemented the above calculations in \texttt{Wolfram Mathematica} \cite{Mathematica} to numerically compute the phase curvature moment $\eta$ and the polarization curvature moment $\xi$ for various optical fibers and optical wavelengths.
Here, we discuss numerical results for telecommunication SMFs and optical nanofibers, and also present an analytical solution to the transport law \eqref{eq:jones transport covariant} for helical optical fibers.

\subsection{Wavelength-dependence in single-mode fibers}

\begin{figure}[t]
	\centering
	\includegraphics[width=\columnwidth]{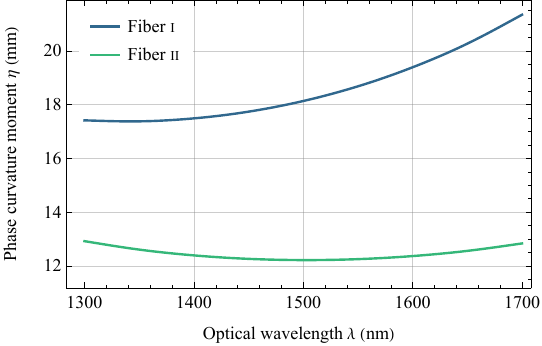}
	\\[2\baselineskip]
	\includegraphics[width=\columnwidth]{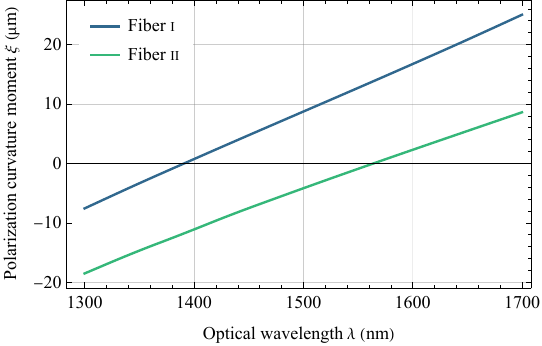}
	\caption{%
		Wavelength-dependence of the phase curvature moment $\eta$ (upper figure) and the polarization curvature moment $\xi$ (lower figure) for typical step-index fibers with a core radius of $\varrho = \SI{4.1}{\micro\meter}$ and refractive indices $n_1 = 1.4712$ and $n_2 = 1.4659$ (Fiber \textsc{i}) as well as $n_1 = 1.4715$ and $n_2 = 1.4648$ (Fiber \textsc{ii}).
	}
	\label{fig:plot:SMF moments}
\end{figure}

\Cref{fig:plot:SMF moments} shows the dependence of the phase curvature moment $\eta$ and the polarization curvature moment $\xi$ on the vacuum wavelength $\lambda = 2 \pi c /\omega$ for two typical telecommunication optical fibers with a core radius $\varrho = \SI{4.1}{\micro\meter}$ and refractive indices (\textsc{i}) $n_1 = 1.4712$ and $n_2 = 1.4659$ as well as (\textsc{ii}) $n_1 = 1.4715$ and $n_2 = 1.4648$.

In his 1987 paper \cite{1987Natur.326..277B}, Berry predicted the polarization curvature moment to be $\xi \approx 1/2\beta = \lambda / (4 \pi \bar n)$ for weakly guiding fibers, where $\bar n$ is the effective refractive index.
Despite the fact that the fibers considered here are weakly guiding (the relative index differences are $\Delta_\textsc{i} = \SI{0.36}{\percent}$ and $\Delta_\textsc{ii} = \SI{0.45}{\percent}$, respectively), we find a deviation from this prediction. Instead of $\xi$ depending \emph{linearly} on $\lambda$, the curves closely follow an \emph{affine} dependence on $\lambda$. The function $\xi(\lambda)$ thus has an isolated zero, at which Rytov’s law extends to second order in perturbation theory.
However, as the following examples show, these curves are not affine throughout.

\subsection{Radius-dependence in nanofibers}

\begin{figure}[t]
	\centering
	\includegraphics[width=\columnwidth]{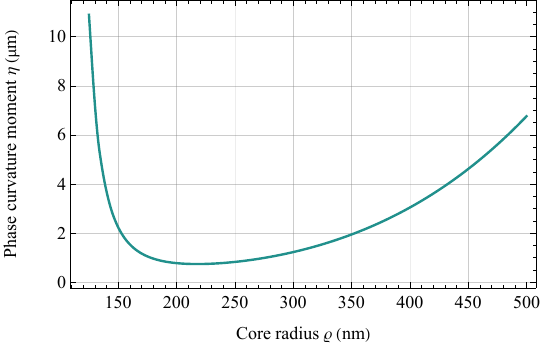}
	\\[2\baselineskip]
	\includegraphics[width=\columnwidth]{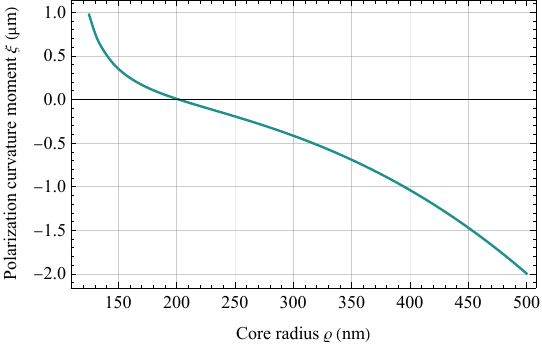}
	\caption{%
		Core radius-dependence of the phase curvature moment $\eta$ (upper figure) and the polarization curvature moment $\xi$ (lower figure) for nanofibers with $n_1 = 1.46$ and $n_2 = 1$, operated at a vacuum wavelength of $\lambda = \SI{633}{\nano\meter}$.
	}
	\label{fig:plot:SMNF moments}
\end{figure}

In recent years, nanofibers, i.e., optical fibers with sub-micrometer diameters, have attracted much experimental interest \cite{2003Natur.426..816T,2014PhRvA..90b3805L,2023PhRvA.107a3713L}. 
In this regime, the two curvature moments $\eta$ and $\xi$ exhibit a behavior which differs significantly from that in telecommunication SMFs described above.
\Cref{fig:plot:SMNF moments} shows the two curvature moments for nanofibers with $n_1 = 1.46$ (fused silica) and $n_2 = 1$ (vacuum), operated at an optical vacuum wavelength of $\lambda = \SI{633}{\nano\meter}$, as was used experimentally in Ref.~\cite{2003Natur.426..816T}.

As such fibers cease to be single-mode at $\varrho \gtrsim \SI{230}{\nano\meter}$ (where modes with $m = 0$ appear) and even support multiple modes with $m = \pm 1$ (labeled by their radial mode index), this plot shows that the polarization curvature moments are continuous functions at the cutoff frequencies of higher modes.

\subsection{Multi-mode regime}

In the above calculations, we have mainly focused on SMFs, as the single-mode condition guarantees that the matrices $\t\Mm{_{m}}$ with $m \neq \pm 1$ are non-singular.
However, we observe numerically that these matrices are typically also non-singular for fibers which are not single-mode, thus allowing us to extend our calculations into the multi-mode regime. It should be noted, however, that not all higher-order modes are linearly polarized, in which case \cref{eq:jones transport covariant} continues to hold on a formal level, but the physical interpretation of $\bs\jones$ becomes less intuitive (since it merely describes the relative weights of the solutions with $m = +1$ and $m = -1$, without a direct relation to a polarization vector).

\begin{figure}[t]
	\centering
	\includegraphics[width=\columnwidth]{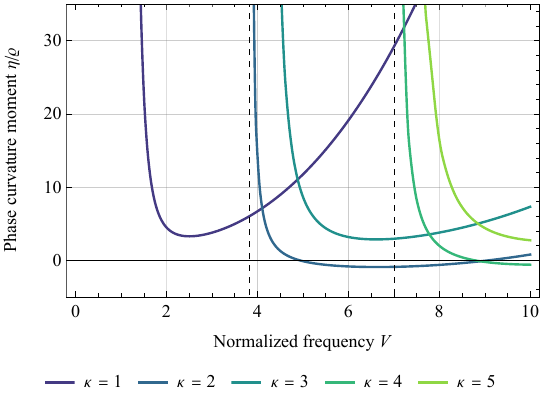}
	\\[2\baselineskip]
	\includegraphics[width=\columnwidth]{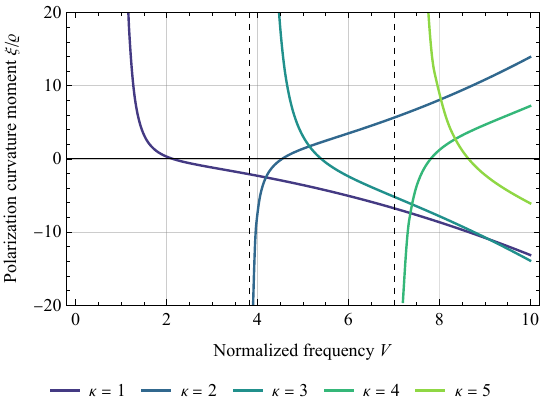}
	\caption{%
		Dependence of the two polarization curvature moments on the normalized frequency $V$ for fibers with $n_1 = 1.46$ and $n_2 = 1$.
		The vertical dashed lines indicate the cutoff frequencies of the higher modes.
	}
	\label{fig:plot:MMNF moments}
\end{figure}

To investigate the behavior of the curvature moments in the multi-mode regime, we consider optical fibers with $n_1 = 1.46$ and $n_2 = 1$ (as in the preceding section) with the normalized frequency $V = \varrho \omega c^{-1} \sqrt{n_1^2 - n_2^2}$ up to $V = 10$, where such fibers support five modes with azimuthal mode index $m = \pm 1$ (they support further modes for different mode indices, but they are not covered by the calculations presented above).
As \cref{fig:plot:MMNF moments} shows, the curvature moments grow steeply near the cutoff frequencies, which are marked by vertical dashed lines.
This indicates a limit on the range of validity of our perturbative scheme, as first-order perturbations of the form $\varepsilon \xi$ and $\varepsilon \eta$ cease to be small near the cutoff.

\subsection{Helical optical fibers}

To illustrate the polarization dynamics derived above, we consider the example of an optical fiber whose baseline forms a helix. In this case, the Fermi--Walker transport for the orthonormal frame \eqref{eq:FW derivative}, as well as the polarization transport equation for the Jones vector \eqref{eq:jones transport covariant}, can be solved exactly.

We start by defining the curve $\gamma$, representing the baseline of the optical fiber, as
\begin{equation}
	\gamma(s) = \left( R \cos (\alpha s), R \sin (\alpha s), H \alpha s \right)\,,
\end{equation}
where $s$ is the arc length,  $R$ is the radius of the helix, $2 \pi H$ is the increase in height of the helix along the $z$-axis after one turn in the $xy$-plane, and $\alpha = (R^2 + H^2)^{-1/2}$. The curve $\gamma$ has constant curvature and torsion
\begin{align}
	\curvature &= R \alpha^2\,,
	&
	\tau &= H \alpha^2\,,
\end{align}
and the Fermi--Walker transported orthonormal frame $(\tangent, e_1, e_2)$ can be constructed explicitly using \cref{eq:fermi_frame}. This frame is illustrated in \cref{fig:fiber}.

At some reference point $s = 0$ on the fiber, the vectors $e_1$ and $e_2$ can be aligned with the initial orientation of transverse electric field lines of the linearly polarized fiber modes given in \cref{fig:plot:polarization}. Then, the Fermi--Walker transport of $e_1$ and $e_2$ along $\gamma$ describes the lowest-order frequency-independent transport law for the polarization of the electromagnetic field along the optical fiber. This is referred to as Rytov’s law, and has been experimentally observed in Refs. \cite{Ross1984,1986PhRvL..57..937T}.

After traveling along $\ell$ complete loops of the helical fiber, the plane of polarization of a linearly polarized electromagnetic wave is rotated by the angle
\begin{align}
	\theta(L) &= 2 \pi \ell H \alpha\,,
	&
	\ell &\in \mathbf N\,.
\end{align}
This is the angle between the vectors $e_i(s=0)$ and $e_i(s = 2\pi \ell / \alpha)$.

Our results also include higher-order frequency-dependent corrections to the polarization dynamics given by Fermi--Walker transport, which can be viewed as an inverse spin Hall effect of light. All information about the polarization of the electromagnetic wave traveling along the optical fiber, relative to the Fermi--Walker transported frame, can be described by the Jones vector $\bs\jones = (\t\jones{_+}, \t\jones{_-})^\mathrm{T}$. The dynamics of the Jones vector is given by the transport equation \cref{eq:transp_jones}, which, for the helical fiber considered here, takes the form
\begin{equation}
	\frac{\dd}{\dd s} \bs\jones
		= i \kappa^2
		\begin{bmatrix}
				\eta
			&   \tfrac{1}{2} \xi e^{- 2 i \tau s}
			\\
				\tfrac{1}{2} \xi e^{+ 2 i \tau s}
			&   \eta
		\end{bmatrix}
		\bs\jones\,.
\end{equation}
The transport equation can be solved exactly in this case, and the general solution can be written as
\begin{subequations}
	\begin{align}
	J_+ &= e^{- i (k + k_-) s} \left( c_1 + c_2 e^{2 i k s} \right)\,, \\
	J_- &= -\frac{2 e^{-i (k - k_+) s}}{ \kappa^2 \xi} \left[ c_1 (\tau + k) + c_2 (\tau - k ) e^{2 i k s} \right]\,,
	\end{align}
\end{subequations}
where
\begin{equation}
	k = \frac{1}{2} \sqrt{4 \tau^2 + \xi^2 \kappa^4}, \qquad k_\pm = \tau \pm \eta \kappa^2\,,
\end{equation}
and $c_1$ and $c_2$ are integration constants depending on the initial values $\mathbf{J}(0)$.
Explicitly, one has
\begin{equation}
	\begin{bmatrix}
		J_+(s)\\
		J_-(s)
	\end{bmatrix}
	=
	U(s)
	\begin{bmatrix}
		J_+(0)\\
		J_-(0)
	\end{bmatrix}\,,
\end{equation}
with the matrix $U(s)$ being given by
\begin{equation}
\begin{split}
	U(s)
		&=
		e^{i \eta \kappa^2 s}
		\begin{bmatrix}
			e^{- i \tau s} & 0\\
			0 & e^{+ i \tau s}
		\end{bmatrix}
		\\&\times
		\begin{bmatrix}
				\cos(k s) + \frac{i \tau}{k} \sin(k s)
			&   \frac{i \xi \kappa^2}{2 k} \sin(k s)
			\\
				\frac{i \xi \kappa^2}{2 k} \sin(k s)
			&   \cos(k s) + \frac{i \tau}{k} \sin(k s)
		\end{bmatrix}\,.
\end{split}
\end{equation}

For illustration, consider an electromagnetic wave that is initially right-hand circularly polarized, thus ${\mathbf{J}(0) = (1,0)^\transpose}$. Then, using the exact solutions given above, one can calculate the probability of observing a beam of opposite circular polarization to be
\begin{equation}
	P_{+-}(s) = \left(1 - \frac{\tau^2}{ k^2}\right) \sin^2(k s)\,.
\end{equation}
This change of polarization represents an inverse spin Hall effect of light, where the polarization dynamics is controlled by the wavelength and by the bending of the optical fiber. For typical fiber parameters such as $R = \SI{10}{\centi\meter}$, $H = \SI{1}{\milli\meter}$ and $\xi = \SI{10}{\micro\meter}$, the above transition probability oscillates with an amplitude $1-\tau^2/k^2 \approx \SI{2.5e-5}{}$ and a period $\pi k^{-1} \approx \SI{31}{\meter}$. Furthermore, note that $k$ depends on the polarization curvature moment $\xi$, which, for telecommunication fibers, is an affine function of the optical wavelength $\lambda$ (see \cref{fig:plot:SMF moments}), but can have a more general behavior for nanofibers (\cref{fig:plot:SMNF moments}) or for higher-order modes (\cref{fig:plot:MMNF moments}). As $\xi$ controls the magnitude of the inverse spin Hall effect of light, the results in \cref{fig:plot:MMNF moments} suggest that this effect becomes large and potentially observable close to the mode cutoff frequencies or for large values of the normalized frequency $V$.

When the torsion $\tau$ is set to zero, the helix reduces to a circle of curvature $\kappa = R^{-1}$, and the above transition probability becomes
\begin{equation}
	P_{+-}(s) \big|_{\tau = 0} =  \sin^2 ( \half \xi \kappa^2 s )\,.
\end{equation}
Thus, a circularly polarized electromagnetic wave can flip to the opposite state of circular polarization after propagating along a circular fiber. This occurs after the electromagnetic wave has traveled a certain length $s$ along the fiber, depending on the geometry (through $\kappa$), the parameters of the fiber, and the wave frequency ($\xi$ depends on the refractive indices in the core and cladding, on the radius of the core, and on wave frequency). More precisely, we will have $P_{+-}(s) \big|_{\tau = 0} = 1$ for $s = (2i+1)\frac{\pi}{\xi \kappa^2}$, with $i \in \mathbf Z$.
A similar transition probability was also obtained in Ref.~\cite[Eq. 32]{2018PhRvA..97c3843L}. However, our result is different in the sense that $\xi$ is an affine function (at least in the case of the telecommunication fibers presented in \cref{fig:plot:SMF moments}) instead of a linear function of $\lambda$, as was the case in Ref.~\cite{2018PhRvA..97c3843L}. Furthermore, in our case $\xi$ contains additional information about the optical fiber parameters, such as the core radius $\varrho$, and the refractive indices of the core and cladding.

We can also calculate the transition probability from an initially right-handed circularly polarized wave to a linearly polarized wave:
\begin{equation}\label{eq:P+x}
\begin{split} 
	P_{+x}(s) = \frac{1}{2} +  \frac{\xi \kappa^2}{4 k^2} \bigg[& 2 \tau \sin^2(k s) \cos(2 \tau s) \\
	& - k \sin(2 k s) \sin(2 \tau s) \bigg]\,.
\end{split}
\end{equation}
Note that while $P_{+-}(s)$ has a period of $\pi k^{-1}$, $P_{+x}(s)$ involves products of two sinusoidal functions with periods of $\pi k^{-1}$ and $\pi \tau^{-1}$, respectively. For the fiber parameters mentioned above, we have $k \approx \tau$, with $k - \tau \approx \SI{1.2e-6}{\per\meter}$. In this case, $P_{+x}(s)$ will oscillate on short length scales with frequency $k \approx \tau$, and we can write
\begin{equation}
	P_{+x}(s) \approx \frac{1}{2}\left[ 1 -  \frac{\xi \kappa^2}{\tau} \sin^2(\tau s) \right] \,.
\end{equation}
In this approximation, the amplitude of the oscillations of $P_{+x}(s)$ is controlled by the wave frequency and fiber parameters (through $\xi$) and by the geometry (through $\kappa$ and $\tau$), while the frequency of the oscillations of $P_{+x}(s)$ is solely controlled by the geometry (through $\tau$). In particular, a transition from circular to linear polarization is achieved when $ P_{+x}(s) = 1$, although this would require some fine tuning of the geometry, fiber parameters and wave frequency.

In general, there will also be a long-scale modulation of $P_{+-}(s)$ due to the small difference between the two frequencies in \cref{eq:P+x}. However, this additional modulation is only significant on length scales of the order $\pi (k - \tau)^{-1} \approx \SI{2.5e6}{\meter}$ and can thus be neglected.

\section{Discussion}
\label{s:discussion}

We have derived an equation for the transport of the electromagnetic field’s polarization and phase in curved optical fibers by perturbatively solving Maxwell’s equations, based on a multiple-scales approximation scheme.
The response of the electromagnetic field to the bending was found to be characterized by two coupling constants: the polarization curvature moment $\xi$ and the phase curvature moment $\eta$. We showed how the polarization curvature moment $\xi$ leads to an inverse spin Hall effect of light, where the state of polarization of the electromagnetic wave is controlled by the wavelength and the bending of the optical fiber.

The equations for the polarization transport derived in this paper are similar in structure to those derived by Berry \cite{1987Natur.326..277B}. In both treatments, the leading-order transport law is given by Fermi--Walker transport, while second-order corrections depend on the fiber’s curvature.
However, the details of the second-order corrections differ: in Berry's result, second-order corrections depend linearly on frequency and quadratically on curvature, but there is no explicit dependence on fiber parameters (neither on the core radius nor on the refractive indices). Furthermore, Berry’s paper does not contain a detailed derivation (the publication of which was announced in the paper, but we are not aware of it).
By contrast, our explicit derivation shows that second-order corrections generally can have a more complicated frequency dependence than predicted by Berry, as described by the phase and polarization curvature moments (which depend on the fiber parameters).

Similar results to Berry’s were also obtained in the related paper by Lai \textit{et al.}~\cite{2018PhRvA..97c3843L}. While our calculations model the electromagnetic field in both the fiber core and the fiber cladding (with appropriate interface conditions), Lai \textit{et al.} evaluated the field equations only up to the fiber core radius (i.e.\ in the core), suggesting that the field was approximated to vanish in the cladding---an approximation which was not necessary in the calculations presented in this paper and is not satisfied in practice for nanofibers.

Moreover, our methods do not rely on the commonly used weak-guidance approximation, nor do they make use of the thin-layer method \cite{2018PhRvA..97c3843L,2019PhRvA.100c3825L}.
Instead, the methods presented here implement a perturbative scheme for the full vectorial Maxwell equations, based only on the assumption of weak and slowly varying bending of the fiber (when compared with the wavelength of the fields therein).

We expect the methods described here to be capable also of providing detailed analyses of fiber mode perturbations by other effects, such as variations in the refractive indices due to stresses, variations in the fiber core radius, but also gravitational effects.

\section*{Acknowledgments}

We are grateful to Lars Andersson, Piotr Chruściel, and Christopher Hilweg for helpful discussions.
T.M. is a recipient of a DOC Fellowship of the Austrian Academy of Sciences at the University of Vienna, Faculty of Physics, and is supported by the Vienna Doctoral School in Physics (VDSP), the research network TURIS, and in part by the Austrian Science Fund (FWF), Project No.~P34274, as well as by the European Union (ERC, GRAVITES, 101071779).

\appendix

\section{Spatial Fermi coordinate system}
\label{s:Fermi coordinates}

This section summarizes the construction of Fermi normal coordinates, based on Fermi--Walker transport.
For an analogous construction based on the Serret--Frenet frame, see, e.g., Refs.~\cite{2003AnPhy.307..132S, 2018PhRvA..97c3843L}.

Let $(M, g)$ be an oriented three-dimensional Riemannian manifold with Levi--Civita connection $\nabla$.
Let $s \mapsto \gamma(s)$ be a curve parameterized by arc length, and denote by $\tangent = \gamma'$ its tangent vector (with $g(\tangent, \tangent) = 1$) and by $\normal = \nabla_\tangent \tangent$ its normal vector.
In regions where the curvature $\kappa = \sqrt{g(\normal, \normal)}$ is non-zero, one can introduce the orthonormal Serret--Frenet frame
\begin{equation}
	\left( \; \tangent, \; \uNormal = \frac{\normal}{\curvature}, \; \binormal = \tangent \times \uNormal \;\right)\,,
\end{equation}
where $\uNormal$ is the unit normal and $\binormal$ is the binormal.
The torsion of the curve $\gamma$ is then defined as
\begin{equation}
	\tau
		= g(\nabla_\tangent \uNormal, \binormal)
		= - g(\nabla_\tangent \binormal, \uNormal)\,,
\end{equation}
and the evolution of the Serret--Frenet frame along the curve is given by the Serret--Frenet equations \cite[Ch.~7.B]{Spivak_v4}
\begin{subequations}
\begin{align}
	\nabla_\tangent \tangent &= \kappa \uNormal\,, \\
	\nabla_\tangent \uNormal &= - \curvature \tangent + \tau \binormal\,, \\
	\nabla_\tangent \binormal &= - \tau \uNormal\,.
\end{align}
\end{subequations}
The Fermi--Walker derivative $\FWD_s$ is defined by its action on any vector $v$ as
\begin{equation}
	\label{eq:FW derivative}
	\FWD_s v
		= \nabla_\tangent v
		+  g(\normal, v)\, \tangent
		- g(\tangent, v)\, \normal\,.
\end{equation}
One readily verifies that $\FWD_s \tangent = 0$, $\FWD_s \uNormal = \tau B$, $\FWD_s \binormal = -\tau \uNormal$ and $\FWD_s g = 0$.

Now, let $(\tangent, e_1, e_2)$ be an orthonormal frame at any “reference point” $p = \gamma(0)$.
Extending $e_1, e_2$ along $\gamma$ by Fermi--Walker transport, i.e.\ $\FWD_s e_1 = \FWD e_2 = 0$, one obtains an orthonormal frame $(\tangent(s), e_1(s), e_2(s))$ along the entire curve $\gamma$. The Fermi--Walker transported vectors $e_1(s)$ and $e_2(s)$ can be explicitly constructed along any curve $\gamma$ with nonvanishing curvature $\curvature$ as
\begin{align}
	\label{eq:fermi_frame}
	\begin{pmatrix}
		e_1(s)\\
		e_2(s)
	\end{pmatrix}
	= \begin{pmatrix}
		\cos \theta(s)
		& - \sin \theta(s)
		\\
		\sin \theta(s)
		& \phantom+ \cos \theta(s)
	\end{pmatrix}
	\begin{pmatrix}
		N(s)\\
		B(s)
	\end{pmatrix}\,,
\end{align}
where
\begin{equation}
	\theta(s) = \int_0^s \torsion(s') \, \dd s' + \theta(0)\,.
\end{equation}
Next, define the function
\begin{equation}
	\label{eq:FW coordinates parametrization}
	\psi(x, y, s)
		= \exp_{\gamma(s)}[x e_1(s) + y e_2(s)]\,,
\end{equation}
where $\exp : TM \to M$ is the exponential map. The implicit function theorem guarantees that the parameters $(s, x, y)$ define a coordinate system in a neighborhood of the curve $\gamma$.

As $(\p_s, \p_x, \p_y) = (\tangent(s), e_1(s), e_2(s))$ is an orthonormal frame along $\gamma$, one finds that along $\gamma$ the metric tensor takes the form
\begin{equation}
	\t g{_i_j} = \t \delta{_i_j}\,.
\end{equation}

The Christoffel symbols along $\gamma$ can be determined using the general formula
\begin{align}
	\t\Gamma{_i_j_k} = g(\t\p{_i}, \nabla_{\t\p{_j}} \t\p{_k})\,.
\end{align}
Using the definition of the Fermi--Walker transport, one finds
\begin{align}
	\begin{pmatrix}
		\nabla_\tangent \tangent\\
		\nabla_\tangent \t e{_1}\\
		\nabla_\tangent \t e{_2}
	\end{pmatrix}
	= \begin{pmatrix}
			0 &
			+ g(\nu,\t e{_1}) &
			+ g(\nu,\t e{_2})
		\\
			- g(\nu, \t e{_1}) &
			0 &
			0
		\\
			- g(\nu, \t e{_2}) &
			0 &
			0
	\end{pmatrix}
	\begin{pmatrix}
		\tangent\\
		\t e{_1}\\
		\t e{_2}
	\end{pmatrix}\,,
\end{align}
from which one can read $\t\Gamma{_i_j_s} = \t\Gamma{_i_s_j}$ and $\t\Gamma{_i_s_s}$.
It thus remains to determine $\t\Gamma{_i_j_k}$ where the last indices are both distinct from $s$.
This is accomplished by considering the curve $\xi \mapsto \psi(s, c_1 \xi, c_2 \xi)$, where $s$, $c_1$ and $c_2$ are constant. By \cref{eq:FW coordinates parametrization}, this is a geodesic, which implies that $c_1 \t\p{_x} + c_2 \t\p{_y}$ is auto-parallel, which means
\begin{equation}
	c_1^2 \nabla_{\t\p{_x}} \t\p{_x}
	+ c_2^2 \nabla_{\t\p{_y}} \t\p{_y}
	+ 2 c_1 c_2 \nabla_{\t\p{_x}} \t\p{_y}
	= 0\,.
\end{equation}
Since the constants $c_1$ and $c_2$ are arbitrary, one finds that all covariant derivatives entering this equation vanish, hence $\t\Gamma{_i_j_k}$ vanishes whenever the last two indices are both distinct from $s$.

In total, one thus finds that the only non-zero Christoffel symbols (up to symmetry) along $\gamma$ are
\begin{subequations}
\begin{align}
	\t\Gamma{_x_s_s} &= + \t\normal{_x}\,,
	&
	\t\Gamma{_y_s_s} &= + \t\normal{_y}\,,
	\\
	\t\Gamma{_s_s_x} &= - \t\normal{_x}\,,
	&
	\t\Gamma{_s_s_y} &= - \t\normal{_y}\,,
\end{align}
\end{subequations}
where $\t\normal{_i} = (0, \t\normal{_x}, \t\normal{_y})$ are the components of the normal vector $\normal$ in the frame $(\tangent, e_1, e_2)$ (they are functions of $s$ alone).
Using $\t\p{_i} \t g{_j_k} = \t\Gamma{_j_k_i} + \t\Gamma{_k_i_j}$, one finds that the only non-zero derivatives of $\t g{_i_j}$ along $\gamma$ are
\begin{align}
	\t\p{_x} \t g{_s_s} &= - 2 \t\normal{_x}\,,
	&
	\t\p{_y} \t g{_s_s} &= - 2 \t\normal{_y}\,,
\end{align}
Expanding the metric along $\gamma$ in a Taylor series in $x$ and $y$, one obtains
\begin{equation}
	\t g{_i_j}
		= \t\delta{_i_j}
		- 2 \t*\delta{_i^s} \t*\delta{_j^s}(\t\normal{_x} x + \t\normal{_y} y)
		+ O(x^2) + O(y^2)\,.
\end{equation}
Up to error terms of order $r^2 = x^2 + y^2$, one thus has
\begin{equation}
\label{eq:metric Fermi coordinates}
	g = \dd x^2 + \dd y^2 + (1 - \t\normal{_i} \t x{^i})^2 \dd s^2 + O(r^2)\,.
\end{equation}
While the calculations leading to \cref{eq:metric Fermi coordinates} were perturbative, it turns out that the error terms vanish in flat Euclidean space. This can be seen by noting that the metric of \cref{eq:metric Fermi coordinates} without the error terms has vanishing Riemann curvature, and that the $xy$-plane is totally geodesic, so that straight coordinate lines in the $xy$-plane are exact solutions to the geodesic equation. Error terms thus arise only in the presence of curvature, and are thus absent for the applications in this paper.

In cylindrical coordinates $(s, r, \vartheta)$, one finds
\begin{equation}
	g =
		\dd r^2 
		+ r^2 \dd \vartheta^2
		+ \left[ 1 - \frac{r}{\sqrt 2} \left(\t\normal{_+} e^{+ i \vartheta} - \t\normal{_-} e^{- i \vartheta} \right) \right]^2 \dd s^2\,,
\end{equation}
where
\begin{equation}
	\t\normal{_\pm} = \tfrac{1}{\sqrt 2}(\t\normal{_x} \mp i \t\normal{_y}) \,.
\end{equation}
In terms of these functions, the curvature $\curvature$ and the torsion $\torsion$ take the form
\begin{align}
	\curvature(s)
		&= \sqrt{2\,\t\normal{_+}(s) \, \t\normal{_-}(s)}\,,
	&
	\torsion
		&= \frac{i}{2} \left(
			\frac{\t*\normal{_+^\prime}(s)}{\t*\normal{_+}(s)}
			- \frac{\t*\normal{_-^\prime}(s)}{\t*\normal{_-}(s)}
		\right)\,.
\end{align}
Conversely, one has
\begin{equation}
	\label{eq:normal complex components}
	\t\normal{_\pm}
		= \frac{\curvature}{\sqrt 2} \exp\left[
			\mp i\left(
				\int_0^s \torsion(s') \dd s'
				+ \text{const.}
			\right)
		\right]\,.
\end{equation}

\section{Inhomogeneities}
\label{s:inhomogeneities}

Here, we give explicit expressions for the inhomogeneities $\t*\inhom{_\mu^{(m,k)}}$ arising in \cref{eq:perturbative system 1,eq:perturbative system 2}.
To express them in a concise way, we introduce some notation.

First, define the two operators
\begin{subequations}
\begin{align}
	\t*{\op L}{_m^k}
		&= \frac{1}{\sqrt 2} \left[
			\frac{\p}{\p r}
			+ \frac{m}{r}
			+ k r^2 \tilde\beta
			\right]
			\,,
	\\
	\t*{\op B}{^\pm}
		&= - \frac{i r \tilde\beta }{\sqrt 2}\left[
			4 \t b{_\pm}
			+ \frac{\p \t b{_\pm}}{\p \varsigma}
		\right]\,.
\end{align}
\end{subequations}
Then, denoting by $\t a{_\mu}$ the column vector of field components defined in \cref{eq:vector a mu}, let $\Op A$, $\Op B$ and $\Op C$ be defined as follows:
\begin{subequations}
\begin{align}
	\t*{\Op A}{_m^+} \t a{_\mu}
		&= \begin{pmatrix}
			\t*{\op L}{_{-m}^2} \t a{_0}\\
			\t*{\op L}{_{-m}^2} \t*a{_\parallel}\\
			\t*{\op L}{_{-(m+1)}^2} \t a{_+}\\
			\t*{\op L}{_{-(m-1)}^2} \t a{_-}\\
		\end{pmatrix}\
		+ 2 i \tilde\beta \t b{_+} 
		\begin{pmatrix}
			0\\
			+ \t a{_+}\\
			0\\
			- \t a{_\parallel}
		\end{pmatrix}\,,
	\\
	\t*{\Op A}{_m^-} \t*a{_\mu}
		&= \begin{pmatrix}
			\t*{\op L}{_{+m}^2} \t*a{_0}\\
			\t*{\op L}{_{+m}^2} \t*a{_\parallel}\\
			\t*{\op L}{_{+(m+1)}^2} \t*a{_+}\\
			\t*{\op L}{_{+(m-1)}^2} \t*a{_-}
		\end{pmatrix}\
		+ 2 i \tilde\beta \t b{_-}
		\begin{pmatrix}
			0 \\
			+ \t a{_-}\\
			- \t a{_\parallel} \\
			0
		\end{pmatrix}\,,
	\\
	\t*{\Op B}{^+} \t a{_\mu}
		&= \t*{\op B}{^+} \t a{_\mu}
		+ [ 2 \t b{_+} \t\p{_\varsigma} + (\t\p{_\varsigma} \t b{_+})] \begin{pmatrix}
			0 \\
			+  \t a{_+} \\
			0 \\
			- \t a{_\parallel}
		\end{pmatrix}\,,
	\\
	\t*{\Op B}{^-} \t a{_\mu}
		&= \t*{\op B}{^-} \t a{_\mu}
		+ [ 2 \t b{_-} \t\p{_\varsigma} + (\t\p{_\varsigma} \t b{_-})]
		\begin{pmatrix}
			0 \\
			+ \t a{_-} \\
			- \t a{_\parallel}\\
			0
		\end{pmatrix}\,,
	\\
	\begin{split}
	\t*{\Op C}{^0} \t a{_\mu}
		&= \sqrt 2 \t b{_+} \t b{_-} r
		\begin{pmatrix}
		   \t*{\op L}{_{0}^3} \t*a{_0}\\
		   \t*{\op L}{_{2}^3} \t*a{_\parallel}\\
		   \t*{\op L}{_{1}^3} \t*a{_+}\\
		   \t*{\op L}{_{1}^3} \t*a{_-}\\
		\end{pmatrix}
		\\&\quad
		+ 2 \sqrt 2 i r \tilde\beta \t b{_+} \t b{_-}
		\begin{pmatrix}
			0 \\
			\t a{_+} + \t a{_-}\\
			- \t a{_\parallel} \\
			- \t a{_\parallel}
		\end{pmatrix}\,,
	\end{split}
	\\
	\begin{split}
		\t*{\Op C}{_m^+} \t a{_\mu}
			&= \frac{r \t*b{_+^2}}{\sqrt 2}
			\begin{pmatrix}
				\t*{\op L}{_{-m}^3} \t*a{_0}\\
				\t*{\op L}{_{-m}^3} \t*a{_\parallel}\\
				\t*{\op L}{_{-(m+1)}^3} \t*a{_+}\\
				\t*{\op L}{_{-(m-1)}^3} \t*a{_-}
			\end{pmatrix}
			\\&\quad
			+ 2 \sqrt 2 i r \tilde\beta \t*b{_+^2}
			\begin{pmatrix}
				0 \\
				\t*a{_+} \\
				0 \\
				- \t*a{_\parallel}
			\end{pmatrix}
			+ \t*b{_+^2}
			\begin{pmatrix}
				0\\    
				0\\    
				0\\    
				\t a{_+}    
			\end{pmatrix}\,,
	\end{split}
	\\
	\begin{split}
		\t*{\Op C}{_m^-} \t a{_\mu}
			&= \frac{r \t*b{_-^2}}{\sqrt 2}
			\begin{pmatrix}
				\t*{\op L}{_{+m}^3} \t*a{_0}\\
				\t*{\op L}{_{+m}^3} \t*a{_\parallel}\\
				\t*{\op L}{_{+(m+1)}^3} \t*a{_+}\\
				\t*{\op L}{_{+(m-1)}^3} \t*a{_-}
			\end{pmatrix}
			\\&\quad
			+ 2 \sqrt 2 i r \tilde\beta \t*b{_-^2}
			\begin{pmatrix}
				0 \\
				+ \t a{_-}\\
				- \t a{_\parallel} \\
				0
			\end{pmatrix}
			+ \t*b{_+^2}
			\begin{pmatrix}
				0\\    
				0\\    
				\t a{_-} \\ 
				0
			\end{pmatrix}\,.
	\end{split}
\end{align}
\end{subequations}
With these definitions at hand, the inhomogeneities of the first-order equations can then be written as
\begin{subequations}
\begin{align}
	\begin{split}
		\t*\inhom{_\mu^{\phantom\pm(0,1)}}
			&= \t*{\Op A}{_{-1}^+} \t*a{_\mu^{(-1,0)}}
			+ \t*{\Op A}{_{+1}^-} \t*a{_\mu^{(+1,0)}}\,,
	\end{split}
	\\
	\begin{split}
		\t*\inhom{_\mu^{(\pm 1,1)}}
			&= - 2 i \tilde\beta \p_\varsigma \t*a{_\mu^{(\pm 1,0)}}\,,
	\end{split}
	\\
	\begin{split}
		\t*\inhom{_\mu^{(\pm 2,1)}}
			&= \t*{\Op A}{_{\pm 1}^\pm} \t*a{_\mu^{(\pm 1,0)}}\,.
	\end{split}
\end{align}
\end{subequations}
Further, the inhomogeneities of the second-order equations are
\begin{subequations}
\begin{align}
	\begin{split}
		\t*\inhom{_\mu^{\phantom\pm(0,2)}}
			&= - 2 i \tilde\beta \t\p{_\varsigma} \t*a{_\mu^{(0,1)}}
			\\&
			+ \t*{\op A}{^+_{-1}} \t*a{_\mu^{(-1,1)}}
			+ \t*{\op A}{^-_{+1}} \t*a{_\mu^{(+1,1)}}
			\\&
			+ \t*{\Op B}{^+} \t*a{_\mu^{(-1,0)}}
			+ \t*{\Op B}{^-} \t*a{_\mu^{(+1,0)}}
			\,,
	\end{split}
	\\
	\begin{split}
		\t*\inhom{_\mu^{(\pm 1,2)}}
			&= - \t*\p{_\varsigma^2} \t*a{_\mu^{(\pm 1,0)}}
			- 2 i \tilde\beta \t*\p{_\varsigma} \t*a{_\mu^{(\pm 1,1)}}
			\\&
			+ \t*{\Op A}{_0^\pm} \t*a{_\mu^{(0,1)}}
			+ \t*{\Op A}{_{\pm 2}^\mp} \t*a{_\mu^{(\pm 2,1)}}
			\\&
			+ \t*{\Op C}{^0} \t*a{_\mu^{(\pm 1,0)}}
			+ \t*{\Op C}{_{\mp 1}^\pm} \t*a{_\mu^{(\mp 1,0)}}
			\,,
	\end{split}
	\\
	\begin{split}
		\t*\inhom{_\mu^{(\pm 2,2)}}
			&= - 2 i \tilde\beta \t\p{_\varsigma} \t*a{_\mu^{(\pm 2,1)}}
			\\&
			+ \t*{\Op A}{_{\pm 1}^\pm} \t*a{_\mu^{(\pm 1,1)}}
			+ \t*{\Op B}{^\pm} \t*a{_\mu^{(\pm 1,0)}}
			\,,
	\end{split}
	\\
	\begin{split}
		\t*\inhom{_\mu^{(\pm 3,2)}}
			&= \t*{\Op A}{_{\pm 2}^\pm} \t*a{_\mu^{(\pm 2,1)}}
			+ \t*{\Op C}{_{\pm 1}^\pm} \t*a{_\mu^{(\pm 1,0)}}
			\,.
	\end{split}
\end{align}    
\end{subequations}
These source terms do not couple the temporal field component $\t a{_0}$ with the spatial components $\t a{_\parallel}$ and $\t a{_\pm}$, which is consistent with the decoupled system given in \cref{eq:field equation abstract}.

\vspace{0.1cm}

\bibliography{bibliography}

\begin{thebibliography}{44}%
\makeatletter
\providecommand \@ifxundefined [1]{%
 \@ifx{#1\undefined}
}%
\providecommand \@ifnum [1]{%
 \ifnum #1\expandafter \@firstoftwo
 \else \expandafter \@secondoftwo
 \fi
}%
\providecommand \@ifx [1]{%
 \ifx #1\expandafter \@firstoftwo
 \else \expandafter \@secondoftwo
 \fi
}%
\providecommand \natexlab [1]{#1}%
\providecommand \enquote  [1]{``#1''}%
\providecommand \bibnamefont  [1]{#1}%
\providecommand \bibfnamefont [1]{#1}%
\providecommand \citenamefont [1]{#1}%
\providecommand \href@noop [0]{\@secondoftwo}%
\providecommand \href [0]{\begingroup \@sanitize@url \@href}%
\providecommand \@href[1]{\@@startlink{#1}\@@href}%
\providecommand \@@href[1]{\endgroup#1\@@endlink}%
\providecommand \@sanitize@url [0]{\catcode `\\12\catcode `\$12\catcode `\&12\catcode `\#12\catcode `\^12\catcode `\_12\catcode `\%12\relax}%
\providecommand \@@startlink[1]{}%
\providecommand \@@endlink[0]{}%
\providecommand \url  [0]{\begingroup\@sanitize@url \@url }%
\providecommand \@url [1]{\endgroup\@href {#1}{\urlprefix }}%
\providecommand \urlprefix  [0]{URL }%
\providecommand \Eprint [0]{\href }%
\providecommand \doibase [0]{https://doi.org/}%
\providecommand \selectlanguage [0]{\@gobble}%
\providecommand \bibinfo  [0]{\@secondoftwo}%
\providecommand \bibfield  [0]{\@secondoftwo}%
\providecommand \translation [1]{[#1]}%
\providecommand \BibitemOpen [0]{}%
\providecommand \bibitemStop [0]{}%
\providecommand \bibitemNoStop [0]{.\EOS\space}%
\providecommand \EOS [0]{\spacefactor3000\relax}%
\providecommand \BibitemShut  [1]{\csname bibitem#1\endcsname}%
\let\auto@bib@innerbib\@empty
\bibitem [{\citenamefont {Senior}\ and\ \citenamefont {Jamro}(2009)}]{senior2009optical}%
  \BibitemOpen
  \bibfield  {author} {\bibinfo {author} {\bibfnamefont {J.~M.}\ \bibnamefont {Senior}}\ and\ \bibinfo {author} {\bibfnamefont {M.~Y.}\ \bibnamefont {Jamro}},\ }\href@noop {} {\emph {\bibinfo {title} {Optical Fiber Communications: Principles And Practice}}}\ (\bibinfo  {publisher} {Pearson Education},\ \bibinfo {year} {2009})\BibitemShut {NoStop}%
\bibitem [{\citenamefont {Lee}(2003)}]{LEE200357}%
  \BibitemOpen
  \bibfield  {author} {\bibinfo {author} {\bibfnamefont {B.}~\bibnamefont {Lee}},\ }\bibfield  {title} {\bibinfo {title} {Review of the present status of optical fiber sensors},\ }\href {https://doi.org/https://doi.org/10.1016/S1068-5200(02)00527-8} {\bibfield  {journal} {\bibinfo  {journal} {Optical Fiber Technology}\ }\textbf {\bibinfo {volume} {9}},\ \bibinfo {pages} {57} (\bibinfo {year} {2003})}\BibitemShut {NoStop}%
\bibitem [{\citenamefont {Predehl}\ \emph {et~al.}(2012)\citenamefont {Predehl}, \citenamefont {Grosche}, \citenamefont {Raupach}, \citenamefont {Droste}, \citenamefont {Terra}, \citenamefont {Alnis}, \citenamefont {Legero}, \citenamefont {Hänsch}, \citenamefont {Udem}, \citenamefont {Holzwarth},\ and\ \citenamefont {Schnatz}}]{doi:10.1126/science.1218442}%
  \BibitemOpen
  \bibfield  {author} {\bibinfo {author} {\bibfnamefont {K.}~\bibnamefont {Predehl}}, \bibinfo {author} {\bibfnamefont {G.}~\bibnamefont {Grosche}}, \bibinfo {author} {\bibfnamefont {S.~M.~F.}\ \bibnamefont {Raupach}}, \bibinfo {author} {\bibfnamefont {S.}~\bibnamefont {Droste}}, \bibinfo {author} {\bibfnamefont {O.}~\bibnamefont {Terra}}, \bibinfo {author} {\bibfnamefont {J.}~\bibnamefont {Alnis}}, \bibinfo {author} {\bibfnamefont {T.}~\bibnamefont {Legero}}, \bibinfo {author} {\bibfnamefont {T.~W.}\ \bibnamefont {Hänsch}}, \bibinfo {author} {\bibfnamefont {T.}~\bibnamefont {Udem}}, \bibinfo {author} {\bibfnamefont {R.}~\bibnamefont {Holzwarth}},\ and\ \bibinfo {author} {\bibfnamefont {H.}~\bibnamefont {Schnatz}},\ }\bibfield  {title} {\bibinfo {title} {A 920-kilometer optical fiber link for frequency metrology at the 19th decimal place},\ }\href {https://doi.org/10.1126/science.1218442} {\bibfield  {journal} {\bibinfo  {journal} {Science}\ }\textbf {\bibinfo {volume} {336}},\ \bibinfo {pages} {441} (\bibinfo {year} {2012})}\BibitemShut {NoStop}%
\bibitem [{\citenamefont {Nayak}\ \emph {et~al.}(2018)\citenamefont {Nayak}, \citenamefont {Sadgrove}, \citenamefont {Yalla}, \citenamefont {Kien},\ and\ \citenamefont {Hakuta}}]{Nayak_2018}%
  \BibitemOpen
  \bibfield  {author} {\bibinfo {author} {\bibfnamefont {K.~P.}\ \bibnamefont {Nayak}}, \bibinfo {author} {\bibfnamefont {M.}~\bibnamefont {Sadgrove}}, \bibinfo {author} {\bibfnamefont {R.}~\bibnamefont {Yalla}}, \bibinfo {author} {\bibfnamefont {F.~L.}\ \bibnamefont {Kien}},\ and\ \bibinfo {author} {\bibfnamefont {K.}~\bibnamefont {Hakuta}},\ }\bibfield  {title} {\bibinfo {title} {Nanofiber quantum photonics},\ }\href {https://doi.org/10.1088/2040-8986/aac35e} {\bibfield  {journal} {\bibinfo  {journal} {Journal of Optics}\ }\textbf {\bibinfo {volume} {20}},\ \bibinfo {pages} {073001} (\bibinfo {year} {2018})}\BibitemShut {NoStop}%
\bibitem [{\citenamefont {Flamini}\ \emph {et~al.}(2018)\citenamefont {Flamini}, \citenamefont {Spagnolo},\ and\ \citenamefont {Sciarrino}}]{Flamini_2019}%
  \BibitemOpen
  \bibfield  {author} {\bibinfo {author} {\bibfnamefont {F.}~\bibnamefont {Flamini}}, \bibinfo {author} {\bibfnamefont {N.}~\bibnamefont {Spagnolo}},\ and\ \bibinfo {author} {\bibfnamefont {F.}~\bibnamefont {Sciarrino}},\ }\bibfield  {title} {\bibinfo {title} {Photonic quantum information processing: a review},\ }\href {https://doi.org/10.1088/1361-6633/aad5b2} {\bibfield  {journal} {\bibinfo  {journal} {Reports on Progress in Physics}\ }\textbf {\bibinfo {volume} {82}},\ \bibinfo {pages} {016001} (\bibinfo {year} {2018})}\BibitemShut {NoStop}%
\bibitem [{\citenamefont {Serafini}\ \emph {et~al.}(2006)\citenamefont {Serafini}, \citenamefont {Mancini},\ and\ \citenamefont {Bose}}]{PhysRevLett.96.010503}%
  \BibitemOpen
  \bibfield  {author} {\bibinfo {author} {\bibfnamefont {A.}~\bibnamefont {Serafini}}, \bibinfo {author} {\bibfnamefont {S.}~\bibnamefont {Mancini}},\ and\ \bibinfo {author} {\bibfnamefont {S.}~\bibnamefont {Bose}},\ }\bibfield  {title} {\bibinfo {title} {Distributed quantum computation via optical fibers},\ }\href {https://doi.org/10.1103/PhysRevLett.96.010503} {\bibfield  {journal} {\bibinfo  {journal} {Physical Review Letters}\ }\textbf {\bibinfo {volume} {96}},\ \bibinfo {pages} {010503} (\bibinfo {year} {2006})}\BibitemShut {NoStop}%
\bibitem [{\citenamefont {Cacciapuoti}\ \emph {et~al.}(2019)\citenamefont {Cacciapuoti}, \citenamefont {Caleffi}, \citenamefont {Tafuri}, \citenamefont {Cataliotti}, \citenamefont {Gherardini},\ and\ \citenamefont {Bianchi}}]{cacciapuoti2019quantum}%
  \BibitemOpen
  \bibfield  {author} {\bibinfo {author} {\bibfnamefont {A.~S.}\ \bibnamefont {Cacciapuoti}}, \bibinfo {author} {\bibfnamefont {M.}~\bibnamefont {Caleffi}}, \bibinfo {author} {\bibfnamefont {F.}~\bibnamefont {Tafuri}}, \bibinfo {author} {\bibfnamefont {F.~S.}\ \bibnamefont {Cataliotti}}, \bibinfo {author} {\bibfnamefont {S.}~\bibnamefont {Gherardini}},\ and\ \bibinfo {author} {\bibfnamefont {G.}~\bibnamefont {Bianchi}},\ }\bibfield  {title} {\bibinfo {title} {Quantum internet: Networking challenges in distributed quantum computing},\ }\href {https://doi.org/10.1109/MNET.001.1900092} {\bibfield  {journal} {\bibinfo  {journal} {IEEE Network}\ }\textbf {\bibinfo {volume} {34}},\ \bibinfo {pages} {137} (\bibinfo {year} {2019})}\BibitemShut {NoStop}%
\bibitem [{\citenamefont {Cacciapuoti}\ \emph {et~al.}(2020)\citenamefont {Cacciapuoti}, \citenamefont {Caleffi}, \citenamefont {Van~Meter},\ and\ \citenamefont {Hanzo}}]{9023997}%
  \BibitemOpen
  \bibfield  {author} {\bibinfo {author} {\bibfnamefont {A.~S.}\ \bibnamefont {Cacciapuoti}}, \bibinfo {author} {\bibfnamefont {M.}~\bibnamefont {Caleffi}}, \bibinfo {author} {\bibfnamefont {R.}~\bibnamefont {Van~Meter}},\ and\ \bibinfo {author} {\bibfnamefont {L.}~\bibnamefont {Hanzo}},\ }\bibfield  {title} {\bibinfo {title} {{When Entanglement Meets Classical Communications: Quantum Teleportation for the Quantum Internet}},\ }\href {https://doi.org/10.1109/TCOMM.2020.2978071} {\bibfield  {journal} {\bibinfo  {journal} {IEEE Transactions on Communications}\ }\textbf {\bibinfo {volume} {68}},\ \bibinfo {pages} {3808} (\bibinfo {year} {2020})}\BibitemShut {NoStop}%
\bibitem [{\citenamefont {{Hilweg}}\ \emph {et~al.}(2017)\citenamefont {{Hilweg}}, \citenamefont {{Massa}}, \citenamefont {{Martynov}}, \citenamefont {{Mavalvala}}, \citenamefont {{Chru{\'s}ciel}},\ and\ \citenamefont {{Walther}}}]{2017NJPh...19c3028H}%
  \BibitemOpen
  \bibfield  {author} {\bibinfo {author} {\bibfnamefont {C.}~\bibnamefont {{Hilweg}}}, \bibinfo {author} {\bibfnamefont {F.}~\bibnamefont {{Massa}}}, \bibinfo {author} {\bibfnamefont {D.}~\bibnamefont {{Martynov}}}, \bibinfo {author} {\bibfnamefont {N.}~\bibnamefont {{Mavalvala}}}, \bibinfo {author} {\bibfnamefont {P.~T.}\ \bibnamefont {{Chru{\'s}ciel}}},\ and\ \bibinfo {author} {\bibfnamefont {P.}~\bibnamefont {{Walther}}},\ }\bibfield  {title} {\bibinfo {title} {{Gravitationally induced phase shift on a single photon}},\ }\href {https://doi.org/10.1088/1367-2630/aa638f} {\bibfield  {journal} {\bibinfo  {journal} {New Journal of Physics}\ }\textbf {\bibinfo {volume} {19}},\ \bibinfo {eid} {033028} (\bibinfo {year} {2017})}\BibitemShut {NoStop}%
\bibitem [{\citenamefont {{Beig}}\ \emph {et~al.}(2018)\citenamefont {{Beig}}, \citenamefont {{Chru{\'s}ciel}}, \citenamefont {{Hilweg}}, \citenamefont {{Kornreich}},\ and\ \citenamefont {{Walther}}}]{2018CQGra..35x4001B}%
  \BibitemOpen
  \bibfield  {author} {\bibinfo {author} {\bibfnamefont {R.}~\bibnamefont {{Beig}}}, \bibinfo {author} {\bibfnamefont {P.~T.}\ \bibnamefont {{Chru{\'s}ciel}}}, \bibinfo {author} {\bibfnamefont {C.}~\bibnamefont {{Hilweg}}}, \bibinfo {author} {\bibfnamefont {P.}~\bibnamefont {{Kornreich}}},\ and\ \bibinfo {author} {\bibfnamefont {P.}~\bibnamefont {{Walther}}},\ }\bibfield  {title} {\bibinfo {title} {{Weakly gravitating isotropic waveguides}},\ }\href {https://doi.org/10.1088/1361-6382/aae873} {\bibfield  {journal} {\bibinfo  {journal} {Classical and Quantum Gravity}\ }\textbf {\bibinfo {volume} {35}},\ \bibinfo {eid} {244001} (\bibinfo {year} {2018})}\BibitemShut {NoStop}%
\bibitem [{\citenamefont {{Mieling}}(2020)}]{2020CQGra..37v5001M}%
  \BibitemOpen
  \bibfield  {author} {\bibinfo {author} {\bibfnamefont {T.~B.}\ \bibnamefont {{Mieling}}},\ }\bibfield  {title} {\bibinfo {title} {{On the influence of Earth's rotation on light propagation in waveguides}},\ }\href {https://doi.org/10.1088/1361-6382/ababb2} {\bibfield  {journal} {\bibinfo  {journal} {Classical and Quantum Gravity}\ }\textbf {\bibinfo {volume} {37}},\ \bibinfo {eid} {225001} (\bibinfo {year} {2020})}\BibitemShut {NoStop}%
\bibitem [{\citenamefont {{Mieling}}(2022)}]{2022PhRvA.106f3511M}%
  \BibitemOpen
  \bibfield  {author} {\bibinfo {author} {\bibfnamefont {T.~B.}\ \bibnamefont {{Mieling}}},\ }\bibfield  {title} {\bibinfo {title} {{Gupta-Bleuler quantization of optical fibers in weak gravitational fields}},\ }\href {https://doi.org/10.1103/PhysRevA.106.063511} {\bibfield  {journal} {\bibinfo  {journal} {Physical Review A}\ }\textbf {\bibinfo {volume} {106}},\ \bibinfo {eid} {063511} (\bibinfo {year} {2022})}\BibitemShut {NoStop}%
\bibitem [{\citenamefont {Liu}(2005)}]{Liu_2005}%
  \BibitemOpen
  \bibfield  {author} {\bibinfo {author} {\bibfnamefont {J.-M.}\ \bibnamefont {Liu}},\ }\href {https://doi.org/10.1017/CBO9780511614255.004} {\emph {\bibinfo {title} {Photonic Devices}}}\ (\bibinfo  {publisher} {Cambridge University Press},\ \bibinfo {year} {2005})\BibitemShut {NoStop}%
\bibitem [{\citenamefont {Davis}(2014)}]{Davis_2014}%
  \BibitemOpen
  \bibfield  {author} {\bibinfo {author} {\bibfnamefont {C.~C.}\ \bibnamefont {Davis}},\ }\href {https://doi.org/10.1017/CBO9781139016629} {\emph {\bibinfo {title} {Lasers and Electro-optics: Fundamentals and Engineering}}},\ \bibinfo {edition} {2nd}\ ed.\ (\bibinfo  {publisher} {Cambridge University Press},\ \bibinfo {year} {2014})\BibitemShut {NoStop}%
\bibitem [{\citenamefont {Okamoto}(2021)}]{okamoto2021fundamentals}%
  \BibitemOpen
  \bibfield  {author} {\bibinfo {author} {\bibfnamefont {K.}~\bibnamefont {Okamoto}},\ }\href {https://doi.org/10.1016/C2017-0-02432-1} {\emph {\bibinfo {title} {Fundamentals of Optical Waveguides}}}\ (\bibinfo  {publisher} {Academic Press},\ \bibinfo {year} {2021})\BibitemShut {NoStop}%
\bibitem [{\citenamefont {Rytov}(1938)}]{rytov}%
  \BibitemOpen
  \bibfield  {author} {\bibinfo {author} {\bibfnamefont {S.~M.}\ \bibnamefont {Rytov}},\ }\bibfield  {title} {\bibinfo {title} {On the transition from wave to geometrical optics},\ }\href@noop {} {\bibfield  {journal} {\bibinfo  {journal} {Doklady Akademii Nauk SSSR}\ }\textbf {\bibinfo {volume} {28}},\ \bibinfo {pages} {263} (\bibinfo {year} {1938})}\BibitemShut {NoStop}%
\bibitem [{\citenamefont {Vladimirskii}(1941)}]{Vladimirskii}%
  \BibitemOpen
  \bibfield  {author} {\bibinfo {author} {\bibfnamefont {V.~V.}\ \bibnamefont {Vladimirskii}},\ }\bibfield  {title} {\bibinfo {title} {On the rotation of the plane of polarization in a curved light ray},\ }\href@noop {} {\bibfield  {journal} {\bibinfo  {journal} {Doklady Akademii Nauk SSSR}\ }\textbf {\bibinfo {volume} {31}},\ \bibinfo {pages} {222} (\bibinfo {year} {1941})}\BibitemShut {NoStop}%
\bibitem [{\citenamefont {Vinitskiĭ}\ \emph {et~al.}(1990)\citenamefont {Vinitskiĭ}, \citenamefont {Derbov}, \citenamefont {Dubovik}, \citenamefont {Markovski},\ and\ \citenamefont {Stepanovskiĭ}}]{Vinitskii_1990}%
  \BibitemOpen
  \bibfield  {author} {\bibinfo {author} {\bibfnamefont {S.~I.}\ \bibnamefont {Vinitskiĭ}}, \bibinfo {author} {\bibfnamefont {V.~L.}\ \bibnamefont {Derbov}}, \bibinfo {author} {\bibfnamefont {V.~M.}\ \bibnamefont {Dubovik}}, \bibinfo {author} {\bibfnamefont {B.~L.}\ \bibnamefont {Markovski}},\ and\ \bibinfo {author} {\bibfnamefont {Y.~P.}\ \bibnamefont {Stepanovskiĭ}},\ }\bibfield  {title} {\bibinfo {title} {Topological phases in quantum mechanics and polarization optics},\ }\href {https://doi.org/10.1070/PU1990v033n06ABEH002598} {\bibfield  {journal} {\bibinfo  {journal} {Soviet Physics Uspekhi}\ }\textbf {\bibinfo {volume} {33}},\ \bibinfo {pages} {403} (\bibinfo {year} {1990})}\BibitemShut {NoStop}%
\bibitem [{\citenamefont {Chru\'sci\'nski}\ and\ \citenamefont {Jamio\l{}kowski}(2012)}]{Chruscinski2012}%
  \BibitemOpen
  \bibfield  {author} {\bibinfo {author} {\bibfnamefont {D.}~\bibnamefont {Chru\'sci\'nski}}\ and\ \bibinfo {author} {\bibfnamefont {A.}~\bibnamefont {Jamio\l{}kowski}},\ }\href {https://doi.org/10.1007/978-0-8176-8176-0} {\emph {\bibinfo {title} {{Geometric phases in classical and quantum mechanics}}}},\ \bibinfo {series} {Progress in Mathematical Physics}, Vol.~\bibinfo {volume} {36}\ (\bibinfo  {publisher} {Birkhäuser},\ \bibinfo {year} {2012})\BibitemShut {NoStop}%
\bibitem [{\citenamefont {Ross}(1984)}]{Ross1984}%
  \BibitemOpen
  \bibfield  {author} {\bibinfo {author} {\bibfnamefont {J.~N.}\ \bibnamefont {Ross}},\ }\bibfield  {title} {\bibinfo {title} {The rotation of the polarization in low birefringence monomode optical fibres due to geometric effects},\ }\href {https://doi.org/10.1007/BF00619638} {\bibfield  {journal} {\bibinfo  {journal} {Optical and Quantum Electronics}\ }\textbf {\bibinfo {volume} {16}},\ \bibinfo {pages} {455} (\bibinfo {year} {1984})}\BibitemShut {NoStop}%
\bibitem [{\citenamefont {{Tomita}}\ and\ \citenamefont {{Chiao}}(1986)}]{1986PhRvL..57..937T}%
  \BibitemOpen
  \bibfield  {author} {\bibinfo {author} {\bibfnamefont {A.}~\bibnamefont {{Tomita}}}\ and\ \bibinfo {author} {\bibfnamefont {R.~Y.}\ \bibnamefont {{Chiao}}},\ }\bibfield  {title} {\bibinfo {title} {{Observation of Berry's topological phase by use of an optical fiber}},\ }\href {https://doi.org/10.1103/PhysRevLett.57.937} {\bibfield  {journal} {\bibinfo  {journal} {Physical Review Letters}\ }\textbf {\bibinfo {volume} {57}},\ \bibinfo {pages} {937} (\bibinfo {year} {1986})}\BibitemShut {NoStop}%
\bibitem [{\citenamefont {Chiao}\ and\ \citenamefont {Wu}(1986)}]{PhysRevLett.57.933}%
  \BibitemOpen
  \bibfield  {author} {\bibinfo {author} {\bibfnamefont {R.~Y.}\ \bibnamefont {Chiao}}\ and\ \bibinfo {author} {\bibfnamefont {Y.-S.}\ \bibnamefont {Wu}},\ }\bibfield  {title} {\bibinfo {title} {{Manifestations of Berry's Topological Phase for the Photon}},\ }\href {https://doi.org/10.1103/PhysRevLett.57.933} {\bibfield  {journal} {\bibinfo  {journal} {Physical Review Letters}\ }\textbf {\bibinfo {volume} {57}},\ \bibinfo {pages} {933} (\bibinfo {year} {1986})}\BibitemShut {NoStop}%
\bibitem [{\citenamefont {{Haldane}}(1986)}]{1986OptL...11..730H}%
  \BibitemOpen
  \bibfield  {author} {\bibinfo {author} {\bibfnamefont {F.~D.~M.}\ \bibnamefont {{Haldane}}},\ }\bibfield  {title} {\bibinfo {title} {{Path dependence of the geometric rotation of polarization in optical fibers}},\ }\href {https://doi.org/10.1364/OL.11.000730} {\bibfield  {journal} {\bibinfo  {journal} {Optics Letters}\ }\textbf {\bibinfo {volume} {11}},\ \bibinfo {pages} {730} (\bibinfo {year} {1986})}\BibitemShut {NoStop}%
\bibitem [{\citenamefont {{Haldane}}(1987)}]{1987PhRvL..59.1788H}%
  \BibitemOpen
  \bibfield  {author} {\bibinfo {author} {\bibfnamefont {F.~D.~M.}\ \bibnamefont {{Haldane}}},\ }\bibfield  {title} {\bibinfo {title} {{Comment on “Observation of Berry's topological phase by use of an optical fiber”}},\ }\href {https://doi.org/10.1103/PhysRevLett.59.1788} {\bibfield  {journal} {\bibinfo  {journal} {Physical Review Letters}\ }\textbf {\bibinfo {volume} {59}},\ \bibinfo {pages} {1788} (\bibinfo {year} {1987})}\BibitemShut {NoStop}%
\bibitem [{\citenamefont {{Berry}}(1987)}]{1987Natur.326..277B}%
  \BibitemOpen
  \bibfield  {author} {\bibinfo {author} {\bibfnamefont {M.~V.}\ \bibnamefont {{Berry}}},\ }\bibfield  {title} {\bibinfo {title} {{Interpreting the anholonomy of coiled light}},\ }\href {https://doi.org/10.1038/326277a0} {\bibfield  {journal} {\bibinfo  {journal} {\nat}\ }\textbf {\bibinfo {volume} {326}},\ \bibinfo {pages} {277} (\bibinfo {year} {1987})}\BibitemShut {NoStop}%
\bibitem [{\citenamefont {{Lai}}\ \emph {et~al.}(2018)\citenamefont {{Lai}}, \citenamefont {{Wang}}, \citenamefont {{Liang}}, \citenamefont {{Wang}},\ and\ \citenamefont {{Zong}}}]{2018PhRvA..97c3843L}%
  \BibitemOpen
  \bibfield  {author} {\bibinfo {author} {\bibfnamefont {M.-Y.}\ \bibnamefont {{Lai}}}, \bibinfo {author} {\bibfnamefont {Y.-L.}\ \bibnamefont {{Wang}}}, \bibinfo {author} {\bibfnamefont {G.-H.}\ \bibnamefont {{Liang}}}, \bibinfo {author} {\bibfnamefont {F.}~\bibnamefont {{Wang}}},\ and\ \bibinfo {author} {\bibfnamefont {H.-S.}\ \bibnamefont {{Zong}}},\ }\bibfield  {title} {\bibinfo {title} {{Electromagnetic wave propagating along a space curve}},\ }\href {https://doi.org/10.1103/PhysRevA.97.033843} {\bibfield  {journal} {\bibinfo  {journal} {Physical Review A}\ }\textbf {\bibinfo {volume} {97}},\ \bibinfo {eid} {033843} (\bibinfo {year} {2018})}\BibitemShut {NoStop}%
\bibitem [{\citenamefont {Onoda}\ \emph {et~al.}(2004)\citenamefont {Onoda}, \citenamefont {Murakami},\ and\ \citenamefont {Nagaosa}}]{Onoda2004}%
  \BibitemOpen
  \bibfield  {author} {\bibinfo {author} {\bibfnamefont {M.}~\bibnamefont {Onoda}}, \bibinfo {author} {\bibfnamefont {S.}~\bibnamefont {Murakami}},\ and\ \bibinfo {author} {\bibfnamefont {N.}~\bibnamefont {Nagaosa}},\ }\bibfield  {title} {\bibinfo {title} {Hall effect of light},\ }\href {https://doi.org/10.1103/PhysRevLett.93.083901} {\bibfield  {journal} {\bibinfo  {journal} {Physical Review Letters}\ }\textbf {\bibinfo {volume} {93}},\ \bibinfo {pages} {083901} (\bibinfo {year} {2004})}\BibitemShut {NoStop}%
\bibitem [{\citenamefont {Duval}\ \emph {et~al.}(2006)\citenamefont {Duval}, \citenamefont {Horv\'ath},\ and\ \citenamefont {Horv\'athy}}]{Duval2006}%
  \BibitemOpen
  \bibfield  {author} {\bibinfo {author} {\bibfnamefont {C.}~\bibnamefont {Duval}}, \bibinfo {author} {\bibfnamefont {Z.}~\bibnamefont {Horv\'ath}},\ and\ \bibinfo {author} {\bibfnamefont {P.~A.}\ \bibnamefont {Horv\'athy}},\ }\bibfield  {title} {\bibinfo {title} {Fermat principle for spinning light},\ }\href {https://doi.org/10.1103/PhysRevD.74.021701} {\bibfield  {journal} {\bibinfo  {journal} {Physical Review D}\ }\textbf {\bibinfo {volume} {74}},\ \bibinfo {pages} {021701} (\bibinfo {year} {2006})}\BibitemShut {NoStop}%
\bibitem [{\citenamefont {Hosten}\ and\ \citenamefont {Kwiat}(2008)}]{Hosten2008}%
  \BibitemOpen
  \bibfield  {author} {\bibinfo {author} {\bibfnamefont {O.}~\bibnamefont {Hosten}}\ and\ \bibinfo {author} {\bibfnamefont {P.}~\bibnamefont {Kwiat}},\ }\bibfield  {title} {\bibinfo {title} {{Observation of the spin Hall effect of light via weak measurements}},\ }\href {https://doi.org/10.1126/science.1152697} {\bibfield  {journal} {\bibinfo  {journal} {Science}\ }\textbf {\bibinfo {volume} {319}},\ \bibinfo {pages} {787} (\bibinfo {year} {2008})}\BibitemShut {NoStop}%
\bibitem [{\citenamefont {Bliokh}\ \emph {et~al.}(2008)\citenamefont {Bliokh}, \citenamefont {Niv}, \citenamefont {Kleiner},\ and\ \citenamefont {Hasman}}]{Bliokh2008}%
  \BibitemOpen
  \bibfield  {author} {\bibinfo {author} {\bibfnamefont {K.~Y.}\ \bibnamefont {Bliokh}}, \bibinfo {author} {\bibfnamefont {A.}~\bibnamefont {Niv}}, \bibinfo {author} {\bibfnamefont {V.}~\bibnamefont {Kleiner}},\ and\ \bibinfo {author} {\bibfnamefont {E.}~\bibnamefont {Hasman}},\ }\bibfield  {title} {\bibinfo {title} {{Geometrodynamics of spinning light}},\ }\href {https://doi.org/10.1038/nphoton.2008.229} {\bibfield  {journal} {\bibinfo  {journal} {Nature Photonics}\ }\textbf {\bibinfo {volume} {2}},\ \bibinfo {pages} {748} (\bibinfo {year} {2008})}\BibitemShut {NoStop}%
\bibitem [{\citenamefont {Bliokh}(2009)}]{Bliokh2009}%
  \BibitemOpen
  \bibfield  {author} {\bibinfo {author} {\bibfnamefont {K.~Y.}\ \bibnamefont {Bliokh}},\ }\bibfield  {title} {\bibinfo {title} {{Geometrodynamics of polarized light: Berry phase and spin Hall effect in a gradient-index medium}},\ }\href {https://doi.org/10.1088/1464-4258/11/9/094009} {\bibfield  {journal} {\bibinfo  {journal} {Journal of Optics A: Pure and Applied Optics}\ }\textbf {\bibinfo {volume} {11}},\ \bibinfo {pages} {094009} (\bibinfo {year} {2009})}\BibitemShut {NoStop}%
\bibitem [{\citenamefont {Bliokh}\ \emph {et~al.}(2015)\citenamefont {Bliokh}, \citenamefont {Rodr\'{i}guez-Fortu\~{n}o}, \citenamefont {Nori},\ and\ \citenamefont {Zayats}}]{SOI_review}%
  \BibitemOpen
  \bibfield  {author} {\bibinfo {author} {\bibfnamefont {K.~Y.}\ \bibnamefont {Bliokh}}, \bibinfo {author} {\bibfnamefont {F.~J.}\ \bibnamefont {Rodr\'{i}guez-Fortu\~{n}o}}, \bibinfo {author} {\bibfnamefont {F.}~\bibnamefont {Nori}},\ and\ \bibinfo {author} {\bibfnamefont {A.~V.}\ \bibnamefont {Zayats}},\ }\bibfield  {title} {\bibinfo {title} {{Spin-orbit interactions of light}},\ }\href {https://doi.org/10.1038/nphoton.2015.201} {\bibfield  {journal} {\bibinfo  {journal} {Nature Photonics}\ }\textbf {\bibinfo {volume} {9}},\ \bibinfo {pages} {796} (\bibinfo {year} {2015})}\BibitemShut {NoStop}%
\bibitem [{\citenamefont {O'Connor}\ \emph {et~al.}(2014)\citenamefont {O'Connor}, \citenamefont {Ginzburg}, \citenamefont {Rodr{\'\i}guez-Fortu{\~n}o}, \citenamefont {Wurtz},\ and\ \citenamefont {Zayats}}]{10.1038/ncomms6327}%
  \BibitemOpen
  \bibfield  {author} {\bibinfo {author} {\bibfnamefont {D.}~\bibnamefont {O'Connor}}, \bibinfo {author} {\bibfnamefont {P.}~\bibnamefont {Ginzburg}}, \bibinfo {author} {\bibfnamefont {F.~J.}\ \bibnamefont {Rodr{\'\i}guez-Fortu{\~n}o}}, \bibinfo {author} {\bibfnamefont {G.~A.}\ \bibnamefont {Wurtz}},\ and\ \bibinfo {author} {\bibfnamefont {A.~V.}\ \bibnamefont {Zayats}},\ }\bibfield  {title} {\bibinfo {title} {Spin--orbit coupling in surface plasmon scattering by nanostructures},\ }\href {https://doi.org/10.1038/ncomms6327} {\bibfield  {journal} {\bibinfo  {journal} {Nature Communications}\ }\textbf {\bibinfo {volume} {5}},\ \bibinfo {pages} {5327} (\bibinfo {year} {2014})}\BibitemShut {NoStop}%
\bibitem [{\citenamefont {Nayak}\ \emph {et~al.}(2022)\citenamefont {Nayak}, \citenamefont {Suchiang}, \citenamefont {Ray}, \citenamefont {Banerjee}, \citenamefont {Gupta},\ and\ \citenamefont {Ghosh}}]{nayak2022momentum}%
  \BibitemOpen
  \bibfield  {author} {\bibinfo {author} {\bibfnamefont {J.~K.}\ \bibnamefont {Nayak}}, \bibinfo {author} {\bibfnamefont {H.}~\bibnamefont {Suchiang}}, \bibinfo {author} {\bibfnamefont {S.~K.}\ \bibnamefont {Ray}}, \bibinfo {author} {\bibfnamefont {A.}~\bibnamefont {Banerjee}}, \bibinfo {author} {\bibfnamefont {S.~D.}\ \bibnamefont {Gupta}},\ and\ \bibinfo {author} {\bibfnamefont {N.}~\bibnamefont {Ghosh}},\ }\bibfield  {title} {\bibinfo {title} {{Momentum domain polarization probing of forward and inverse spin Hall effect of leaky modes in plasmonic crystals}},\ }\href {https://arxiv.org/abs/2204.03699} {\bibfield  {journal} {\bibinfo  {journal} {arXiv:2204.03699}\ } (\bibinfo {year} {2022})}\BibitemShut {NoStop}%
\bibitem [{\citenamefont {{Gordon}}(1923)}]{1923AnP...377..421G}%
  \BibitemOpen
  \bibfield  {author} {\bibinfo {author} {\bibfnamefont {W.}~\bibnamefont {{Gordon}}},\ }\bibfield  {title} {\bibinfo {title} {{Zur Lichtfortpflanzung nach der Relativit{\"a}tstheorie}},\ }\href {https://doi.org/10.1002/andp.19233772202} {\bibfield  {journal} {\bibinfo  {journal} {Annalen der Physik}\ }\textbf {\bibinfo {volume} {377}},\ \bibinfo {pages} {421} (\bibinfo {year} {1923})}\BibitemShut {NoStop}%
\bibitem [{\citenamefont {{Bender}}\ and\ \citenamefont {{Orszag}}(1978)}]{1978amms.book.....B}%
  \BibitemOpen
  \bibfield  {author} {\bibinfo {author} {\bibfnamefont {C.~M.}\ \bibnamefont {{Bender}}}\ and\ \bibinfo {author} {\bibfnamefont {S.~A.}\ \bibnamefont {{Orszag}}},\ }\href {https://doi.org/10.1007/978-1-4757-3069-2} {\emph {\bibinfo {title} {{Advanced Mathematical Methods for Scientists and Engineers}}}},\ edited by\ \bibinfo {editor} {\bibfnamefont {R.}~\bibnamefont {Ciofalo}},\ International series in pure and applied mathematics\ (\bibinfo  {publisher} {McGraw-Hill},\ \bibinfo {year} {1978})\BibitemShut {NoStop}%
\bibitem [{\citenamefont {Korenev}(2002)}]{korenev2002bessel}%
  \BibitemOpen
  \bibfield  {author} {\bibinfo {author} {\bibfnamefont {B.~G.}\ \bibnamefont {Korenev}},\ }\href@noop {} {\emph {\bibinfo {title} {Bessel functions and their applications}}}\ (\bibinfo  {publisher} {Taylor and Francis},\ \bibinfo {address} {London},\ \bibinfo {year} {2002})\BibitemShut {NoStop}%
\bibitem [{\citenamefont {{Wolfram Research, Inc.}}(2021)}]{Mathematica}%
  \BibitemOpen
  \bibfield  {author} {\bibinfo {author} {\bibnamefont {{Wolfram Research, Inc.}}},\ }\href {https://www.wolfram.com/mathematica} {\bibinfo {title} {Mathematica, {V}ersion 13.0.0}} (\bibinfo {year} {2021}),\ \bibinfo {note} {{Champaign, US-IL}, 2021}\BibitemShut {NoStop}%
\bibitem [{\citenamefont {{Tong}}\ \emph {et~al.}(2003)\citenamefont {{Tong}}, \citenamefont {{Gattass}}, \citenamefont {{Ashcom}}, \citenamefont {{He}}, \citenamefont {{Lou}}, \citenamefont {{Shen}}, \citenamefont {{Maxwell}},\ and\ \citenamefont {{Mazur}}}]{2003Natur.426..816T}%
  \BibitemOpen
  \bibfield  {author} {\bibinfo {author} {\bibfnamefont {L.}~\bibnamefont {{Tong}}}, \bibinfo {author} {\bibfnamefont {R.~R.}\ \bibnamefont {{Gattass}}}, \bibinfo {author} {\bibfnamefont {J.~B.}\ \bibnamefont {{Ashcom}}}, \bibinfo {author} {\bibfnamefont {S.}~\bibnamefont {{He}}}, \bibinfo {author} {\bibfnamefont {J.}~\bibnamefont {{Lou}}}, \bibinfo {author} {\bibfnamefont {M.}~\bibnamefont {{Shen}}}, \bibinfo {author} {\bibfnamefont {I.}~\bibnamefont {{Maxwell}}},\ and\ \bibinfo {author} {\bibfnamefont {E.}~\bibnamefont {{Mazur}}},\ }\bibfield  {title} {\bibinfo {title} {{Subwavelength-diameter silica wires for low-loss optical wave guiding}},\ }\href {https://doi.org/10.1038/nature02193} {\bibfield  {journal} {\bibinfo  {journal} {\nat}\ }\textbf {\bibinfo {volume} {426}},\ \bibinfo {pages} {816} (\bibinfo {year} {2003})}\BibitemShut {NoStop}%
\bibitem [{\citenamefont {{Le Kien}}\ and\ \citenamefont {{Rauschenbeutel}}(2014)}]{2014PhRvA..90b3805L}%
  \BibitemOpen
  \bibfield  {author} {\bibinfo {author} {\bibfnamefont {F.}~\bibnamefont {{Le Kien}}}\ and\ \bibinfo {author} {\bibfnamefont {A.}~\bibnamefont {{Rauschenbeutel}}},\ }\bibfield  {title} {\bibinfo {title} {{Anisotropy in scattering of light from an atom into the guided modes of a nanofiber}},\ }\href {https://doi.org/10.1103/PhysRevA.90.023805} {\bibfield  {journal} {\bibinfo  {journal} {Physical Review A}\ }\textbf {\bibinfo {volume} {90}},\ \bibinfo {eid} {023805} (\bibinfo {year} {2014})}\BibitemShut {NoStop}%
\bibitem [{\citenamefont {{Le Kien}}\ \emph {et~al.}(2023)\citenamefont {{Le Kien}}, \citenamefont {{Nic Chormaic}},\ and\ \citenamefont {{Busch}}}]{2023PhRvA.107a3713L}%
  \BibitemOpen
  \bibfield  {author} {\bibinfo {author} {\bibfnamefont {F.}~\bibnamefont {{Le Kien}}}, \bibinfo {author} {\bibfnamefont {S.}~\bibnamefont {{Nic Chormaic}}},\ and\ \bibinfo {author} {\bibfnamefont {T.}~\bibnamefont {{Busch}}},\ }\bibfield  {title} {\bibinfo {title} {{Direction-dependent coupling between a nanofiber-guided light field and a two-level atom with an electric quadrupole transition}},\ }\href {https://doi.org/10.1103/PhysRevA.107.013713} {\bibfield  {journal} {\bibinfo  {journal} {Physical Review A}\ }\textbf {\bibinfo {volume} {107}},\ \bibinfo {eid} {013713} (\bibinfo {year} {2023})}\BibitemShut {NoStop}%
\bibitem [{\citenamefont {{Lai}}\ \emph {et~al.}(2019)\citenamefont {{Lai}}, \citenamefont {{Wang}}, \citenamefont {{Liang}},\ and\ \citenamefont {{Zong}}}]{2019PhRvA.100c3825L}%
  \BibitemOpen
  \bibfield  {author} {\bibinfo {author} {\bibfnamefont {M.-Y.}\ \bibnamefont {{Lai}}}, \bibinfo {author} {\bibfnamefont {Y.-L.}\ \bibnamefont {{Wang}}}, \bibinfo {author} {\bibfnamefont {G.-H.}\ \bibnamefont {{Liang}}},\ and\ \bibinfo {author} {\bibfnamefont {H.-S.}\ \bibnamefont {{Zong}}},\ }\bibfield  {title} {\bibinfo {title} {{Geometrical phase and Hall effect associated with the transverse spin of light}},\ }\href {https://doi.org/10.1103/PhysRevA.100.033825} {\bibfield  {journal} {\bibinfo  {journal} {Physical Review A}\ }\textbf {\bibinfo {volume} {100}},\ \bibinfo {eid} {033825} (\bibinfo {year} {2019})}\BibitemShut {NoStop}%
\bibitem [{\citenamefont {{Schuster}}\ and\ \citenamefont {{Jaffe}}(2003)}]{2003AnPhy.307..132S}%
  \BibitemOpen
  \bibfield  {author} {\bibinfo {author} {\bibfnamefont {P.~C.}\ \bibnamefont {{Schuster}}}\ and\ \bibinfo {author} {\bibfnamefont {R.~L.}\ \bibnamefont {{Jaffe}}},\ }\bibfield  {title} {\bibinfo {title} {{Quantum mechanics on manifolds embedded in Euclidean space}},\ }\href {https://doi.org/10.1016/S0003-4916(03)00080-0} {\bibfield  {journal} {\bibinfo  {journal} {Annals of Physics}\ }\textbf {\bibinfo {volume} {307}},\ \bibinfo {pages} {132} (\bibinfo {year} {2003})}\BibitemShut {NoStop}%
\bibitem [{\citenamefont {Spivak}(1999)}]{Spivak_v4}%
  \BibitemOpen
  \bibfield  {author} {\bibinfo {author} {\bibfnamefont {M.}~\bibnamefont {Spivak}},\ }\href@noop {} {\emph {\bibinfo {title} {{A Comprehensive Introduction to Differential Geometry}}}},\ Vol.~\bibinfo {volume} {4}\ (\bibinfo  {publisher} {Publish or Perish, Inc.},\ \bibinfo {year} {1999})\BibitemShut {NoStop}%
\end{thebibliography}%
\end{document}